\documentclass[preprint]{aastex}

\newcommand{\be}{\begin{equation}}
\newcommand{\ee}{\end{equation}}
\newcommand{\tturb}{{\tau_T}} 
\newcommand{\tturbk}{{ \tau_{Tk} }} 
\newcommand{\xmax}{{ x_{\rm max} }} 
\newcommand{\diffcon}{{ \beta }} 
\newcommand{\gambar}{{ \langle \gamma \rangle }} 
\newcommand{\mearth}{{ M_\earth }} 
\newcommand{\rdisk}{{ r_{\rm disk} }}
\newcommand{\jdisk}{{ j_{\rm disk} }}
\newcommand{\qm}{{ Q_{\rm m} }}

\begin{document} 

\title{\bf GENERAL ANALYSIS OF TYPE I PLANETARY MIGRATION \\
WITH STOCHASTIC PERTURBATIONS}  

\author{Fred C. Adams$^{1,2}$ and Anthony M. Bloch$^{1,3}$} 

\affil{$^1$Michigan Center for Theoretical Physics \\
Physics Department, University of Michigan, Ann Arbor, MI 48109} 

\affil{$^2$Astronomy Department, University of Michigan, Ann Arbor, MI 48109} 

\affil{$^3$Department of Mathematics, University of Michigan, 
Ann Arbor, MI 48109} 

\begin{abstract} 

This paper presents a generalized treatment of Type I planetary
migration in the presence of stochastic perturbations. In many
planet-forming disks, the Type I migration mechanism, driven by
asymmetric torques, acts on a short time scale and compromises planet
formation. If the disk also supports MHD instabilities, however, the
corresponding turbulent fluctuations produce additional stochastic
torques that modify the steady inward migration scenario. This work
studies the migration of planetary cores in the presence of stochastic
fluctuations using complementary methods, including a Fokker-Planck
approach and iterative maps. Stochastic torques have two main effects:
[1] Through outward diffusion, a small fraction of the planetary cores
can survive in the face of Type I inward migration. [2] For a given
starting condition, the result of any particular realization of
migration is uncertain, so that results must be described in terms of
the distributions of outcomes.  In addition to exploring different
regimes of parameter space, this paper considers the effects of the
outer disk boundary condition, varying initial conditions, and
time-dependence of the torque parameters. For disks with finite radii,
the fraction of surviving planets decreases exponentially with time.
We find the survival fractions and decay rates for a range of disk
models, and find the expected distribution of locations for surviving
planets. For expected disk properties, the survival fraction lies in
the range $0.01 < p_S < 0.1$.

\end{abstract} 

\keywords{MHD --- planetary systems --- planetary systems: formation ---
planets and satellites: formation --- turbulence} 

\section{INTRODUCTION} 

The past decade has led to tremendous progress in our understanding of
extrasolar planets and the processes involved in planet formation.
These advances include both observations, which now include the
detection of nearly 300 planets outside our Solar System (see, e.g.,
Udry et al. 2007 for a recent review), along with a great deal of
accompanying theoretical work. One surprise resulting from the
observations is the finding that extrasolar planets display a much
wider range of orbital configurations than was anticipated. Planets
thus move (usually inward) from their birth sites, or while they are
forming, in a process known as planet migration (e.g., see Papaloizou
\& Terquem 2006 for a recent review). 

The migration process is especially rapid when the planets have small
masses, less than $\sim$30 $\mearth$, so they cannot clear gaps in the
disks (Goldreich \& Tremaine 1979, 1980).  This phase is often called
Type I migration (Ward 1997ab, Tanaka et al. 2002) and can cause a
forming planet to be accreted onto its central star in about 0.1 -- 1
Myr, time scales shorter than the expected time (1 -- 10 Myr) required
for Jovian planets to attain their final masses (e.g., Lissauer \&
Stevenson 2007). However, if a growing planet can attain a mass
greater than $\sim$30 -- 100 $\mearth$ before accretion, it can clear
a gap in the disk, and its subsequent migration rate is much smaller
(this latter process is known as Type II migration).  We note that the
mass required for gap clearing depends on viscosity, scale height, and
other disk parameters, so that a range of values is expected (for
further detail, see Ward 1997a, especially Figure 14). In any case,
the forming planet must grow massive enough --- quickly enough --- in
order to survive.  The problem is made more urgent because the Type I
migrate rate increases with increasing planetary mass until the
gap-clearing threshold is reached.  This dilemma is generally known as
the ``Type I Migration Problem'' and can be alleviated by the action
of stochastic torques produced by disk turbulence. These torques drive
random walk behavior that allows some fraction of the growing
planetary cores to survive.  The goal of this paper is to study Type I
migration in the presence of stochastic torques in order to assess the
expected survival rates for forming planets and to elucidate the
physics of this mechanism.

A significant body of previous work exists. Initial explorations of
the effects of turbulence on Type I migration showed that stochastic
torques can dominate the steady inward torques and thus have the
potential to allow more planets to survive (Laughlin et al. 2004,
hereafter LSA; Nelson \& Papaloizou 2004, hereafter NP).  Subsequent
numerical studies demonstrated the corresponding random walk behavior
of the migrating planets and explored the possible range of turbulent
fluctuation amplitudes and correlation times (Nelson 2005, Papaloizou
et al. 2007). Due to computational limitations, however, full
numerical simulations that simultaneously include MHD turbulence and
planetary migration can only be carried out for hundreds of orbits,
whereas the expected time scale of interest is millions of years (and
hence millions of orbits). As a result, long term behavior must be
studied using analytical and statistical methods. Preliminary results
were given in LSA, and then a more comprehensive treatment using
Fokker-Planck methods was developed (Johnson et al. 2006, hereafter
JGM). This latter work showed that only a small fraction of the planet
population is expected to survive in the long term, and also
considered the effects of disk structure on the results (e.g.,
departures from power-law surface density and temperature profiles ---
see also Menou \& Goodman 2004).  This present paper also adopts an
analytical/statistical approach in order to study the long-term
outcome of this migration mechanism.  Our goal is thus to generalize
the previous analyses of LSA and JGM.

This paper extends previous work in several ways: We explore the
effects of the outer boundary condition. In particular, if the disk
has an outer edge, as expected for young star/disk systems (typically
with $\rdisk \sim 30 - 100$ AU), the outer boundary condition affects
the dynamics by enforcing exponential decay in the number of surviving
planets. In contrast, the survival fraction decays as a power-law
decay in the limit where $\rdisk \to \infty$. We also consider the
effects of the initial conditions on the survival rates; planets
formed in the outer disk have a much greater chance of survival,
compared with those formed in the inner disk, with the boundary close
to $r \sim 10$ AU (near the expected locations for planetary cores to
form).  Next we consider the possible effects of time dependence on
the migration torques.  Over the time span of interest, millions of
years, the disk mass and the disk surface density decrease with time,
whereas the mass of the migrating planetary core will grow.  Both of
these effects lead to time varying torque parameters, which are
modeled herein.  In the long time limit, we find the distribution of
surviving planets by solving for the lowest-order eigenfunction of the
Fokker-Planck equation.  The dynamics of this migration problem are
surprisingly rich. For example, although turbulent torques lead to
random walk behavior and allow planets to survive, large amplitude
fluctuations actually {\it reduce} the survival fraction; we explore
the interplay between these competing outcomes and solve the
corresponding optimization problem.  Finally, we present an iterative
map approach. In addition to providing an alternate description for
the dynamics of the migration problem, this approach easily allows for
the inclusion of eccentricity variations and large fluctuations.

This paper is organized as follows. We present our formulation of the
Type I migration torques and turbulent forcing in Section 2.  Section
3 develops a Fokker-Planck approach to the dynamics, including the
basic formulation, analytic results for the cases where inward
migration and diffusion are considered in isolation, as well as a
self-similar model. Numerical solutions to the Fokker-Planck equation
are presented in Section 4, which contains the main astronomical
results (outlined above).  The paper concludes in Section 5 with a
summary of our results and a discussion of their implications.  The
Appendix presents an alternate approach to the migration problem using
an iterative mapping scheme; this treatment not only adds to our
understanding of the underlying dynamics, it can also be used to
include larger stochastic perturbations, different boundary
conditions, and additional variables.

\section{FORMULATION} 

\subsection{Basic Disk Properties} 

In order to explore the wide range of possible effects that arise in
this coupled migration problem, we consider simple power-law disk
models. Specifically, the surface density and temperature distribution
of the disks are taken to be power-laws in radius,
\be
\Sigma (r) = \Sigma_1 \left( {r_1 \over r} \right)^p 
\qquad {\rm and} \qquad 
T (r) = T_1 \left( {r_1 \over r} \right)^q \, . 
\label{powerlaw} 
\ee
The normalization constants are determined by the total disk mass and total
effective disk luminosity, respectively.  In this formulation, we take
$r_1$ = 1 AU, so the coefficients $\Sigma_1$ and $T_1$ correspond to
their values at 1 AU. The index $p$ is expected to lie in the range
$p$ = 1 -- 2, with a typical value $p$ = 3/2. This latter value arises
from the Minimum Mass Solar Nebula (e.g., Weidenschilling 1977).
Considerations of disk formation during protostellar collapse produce
indices in the range $p$ = 3/2 -- 7/4 (Cassen \& Moosman 1981, Adams
\& Shu 1986).  The normalization for the surface density has a
benchmark value of $\Sigma_1 \approx 4500$ g/cm$^2$ (e.g., Kuchner
2004, Weidenschilling 1977). The power law index of the temperature 
profile is expected to be $q \approx 3/4$ for a viscous accretion disk
(e.g. Pringle 1981) and a flat reprocessing disk (Adams \& Shu 1986), 
whereas $q \approx 1/2$ for a flared reprocessing disk (Chiang \&
Goldreich 1997). The latter value also applies to the early solar 
nebula (Weidenschilling 1977). 

The disk is assumed to be purely Keplerian, and the orbits are taken
to be circular, so that the orbital angular momentum $j$ is given by
\be
j = m_P \left( G M_\ast r \right)^{1/2} \, . 
\ee
Further, the disk scale height $H$ is given by $H = a_S / \Omega$,
where $a_S$ is the sound speed, which is in turn determined by the
disk temperature profile. As shown below, the formulation of this
paper requires specification of the scale height, rather than the
temperature distribution itself, and we adopt the form 
\be
{H \over r} = \left( {H \over r} \right)_1 
\left( {r \over r_1} \right)^{(1-q)/2} \, . 
\label{hover} 
\ee
A benchmark value for the scale height at $r_1$ = 1 AU is
$H/r$ = 0.1.  

\subsection{Turbulent Forcing} 

The net effect of turbulence is to provide stochastic forcing
perturbations.  We first specify the time scale $\tturb$ required for
the disk to produce an independent realization of the turbulent
fluctuations. Previous work (LSA, NP, Nelson 2005) indicates that this
time scale is approximately an orbit time, so we parameterize the time
scale according to
\be
\tturb = f_\alpha {2 \pi \over \Omega} \, , 
\label{tturbdef} 
\ee
where $\Omega$ is the Keplerian rotation rate and where $f_\alpha$ is
a dimensionless parameter of order unity.  Note that this time scale
varies with radial location in the disk.

Next we need to determine the amplitudes $[(\Delta j)/j]_k$ of the
angular momentum perturbations due to turbulent forcing.  In general,
the torque exerted on a planet by the disk will be a fraction of the
benchmark scale $T_D$ given by
\be
T_D  = 2 \pi G \Sigma r m_P \, , 
\label{bench} 
\ee 
where $\Sigma$ is the disk surface density (e.g., JGM). 
The amplitude for angular momentum variations is thus given by 
\be
{\Delta j} = f_T T_D \tturb \, , 
\ee
where $\tturb$ is the time over which one independent realization of
the turbulence acts. The total torque produced by the turbulence is a
fraction $f_T$ of the benchmark scale given by equation (\ref{bench}).
These turbulent forcing amplitudes have been estimated using MHD
simulations (e.g., LSA, NP, Nelson 2005), which show that 
$f_T \sim 0.05$ (with a range of variation about this typical value).  
The relative fluctuation amplitude is then given by
\be 
\left( {\Delta j \over j} \right)_T = f_\alpha f_T (2 \pi)^2 
{\Sigma r^2 \over M_\ast} \, . 
\label{ampturb} 
\ee
With $f_T = 0.05$ and $f_\alpha$ = 1, the leading numerical
coefficient becomes $\pi^2/5 \sim 2$. The expression in equation
(\ref{ampturb}) determines the fluctuation amplitude. The actual
changes in angular momentum over a given time scale $\tturb$ are thus
given by
\be
{\Delta j \over j} = \left( {\Delta j \over j} \right)_T \xi 
= f_\alpha f_T (2 \pi)^2 {\Sigma r^2 \over M_\ast} \, \xi \, , 
\ee
where the random variable $\xi$ has zero mean and unit variance. In 
this work, we assume that $\xi$ has a gaussian distribution. 

Note that this treatment also assumes that the planet is small enough
so that it has no back reaction on the disk. Since we are primarily
interested in planetary cores in the mass range $m_P$ = 1 -- 30
$\mearth$, this assumption is expected to be valid. Planets of larger
mass are likely to clear gaps in their immediate vicinity within the
disk (Goldreich \& Tremaine 1980), however, and hence the turbulent
torques are reduced in such systems (this reduction can be included in
the formalism; see Adams et al. 2008).

For power-law disks, the relative fluctuation amplitude varies 
with radius according to 
\be
\left( {\Delta j \over j} \right)_T \propto r^{2 - p} \, . 
\ee
For a typical value of the power-law index is $p$ = 3/2, the relative
fluctuations $[(\Delta j)/j] \sim r^{1/2} \sim j$.

\subsection{Type I Migration} 

The strength of Type I torques are given by 
\be
T_1 = f_1 \left( {m_P \over M_\ast} \right)^2 
\pi \Sigma r^2 (r \Omega)^2 \left( {r \over H} \right)^2 \, , 
\ee
where $f_1$ is a dimensionless (constant) parameter (Ward 1997a).  
Over the same time scale $\tturb$ used to evaluate the changes in
angular momentum due to turbulence, the corresponding changes due 
to Type I torques are given by
\be
\left( {\Delta j \over j} \right)_1 = 
f_1 f_\alpha 2 \pi^2 \left( {m_P \over M_\ast} \right)
{\Sigma r^2 \over M_\ast} \left( {r \over H} \right)^2 \, . 
\ee 
For power-law disks, the Type I angular momentum increments 
vary with radius according to 
\be
\left( {\Delta j \over j} \right)_1 \propto r^{1 + q - p} \, .
\ee 
For typical indices $p$ = 3/2 and $q$ = 3/4, the Type I angular
momentum increments vary relatively slowly with radius, i.e.,
$[(\Delta j)/j] \sim r^{1/4} \sim j^{1/2}$. For the particular values
$p$ = 3/2 and $q$ = 1/2, often used to model the early solar nebula,
the relative fluctuation $[(\Delta j)/j]$ is a constant with respect
to radius $r$.  For typical values of the input parameters, the
constant amplitude of the angular momentum increment is given by 
$[(\Delta j)/j] \sim 10^{-5}$. 

\subsection{Comparison of Time Scales}

The Type I migration torques provide a steady inward forcing on the
planets, whereas the turbulent torques are stochastic. At a given
radial location in the disk, or equivalently at a given value of
angular momentum $j$, the ratio of the time scales for the two types
of torques to move the planet is given by
\be
{t_1 \over t_T} = { [(\Delta j)/j]_T^2 \over [(\Delta j)/j]_1 } = 
{8 \pi^2 f_\alpha f_T^2 \over f_1} 
\left( {\Sigma H^2 \over m_P } \right) \, . 
\label{timeratio} 
\ee 
The expected value of the leading coefficient is $\sim$ 1/5.  
For power-law disks, this ratio of time scales varies with 
radius according to 
\be
{t_1 \over t_T} \propto r^{3 - p - q} \, . 
\ee
As a result, the time scale ratio grows (approximately) linearly with
radius. More significantly, the power-law index is always positive,
even for the most extreme parameters expected in planet-forming disks,
so that the outer disk is dominated by turbulent migration, while the
inner disk is dominated by Type I migration. 

The above discussion motivates the definition of a dimensionless
parameter $\qm$ that determines the characteristics of planetary
migration at a given radial location in the disk:
\be
\qm \equiv {8 \pi^2 f_T^2 \Sigma H^2 \over m_P} \, , 
\label{qdef} 
\ee
where we have ignored the parameters $f_1$ and $f_\alpha$ since they
are expected to be close to unity.  For $\qm > 1$, turbulent torques
dominate and migration behaves as a random walk. For $\qm < 1$, Type I
torques dominate and planets migrate steadily inward. For typical disk
parameters, we expect $\qm \sim 0.1$ near $r$ = 1 AU. Keep in mind
that the value of $\qm$ depends on both the radial location in the
disk and on time. As the disk and planet evolve, the surface density
$\Sigma$ grows smaller, while the planetary core mass $m_P$ grows
larger, so that $\qm$ is generally a decreasing function of time.

On a related note, we can estimate the time required for the two
migration mechanisms to move planets from a given starting point in
the disk to either the inner or outer disk edge. The time required for
Type I migration to move a planet inward to the star from a starting
angular momentum value $j_0$ is approximately given by $t_I \approx
j_0 / [3 T_1 (j_0)]$. To fix ideas, we take the starting radius to be
10 AU. For typical values of the torque parameters, the Type I time
scale for inward migration is $t_I \sim 1$ Myr.  For comparison, we
can estimate the time required for diffusion to transport planets to
the outer disk edge. The timescale for the distribution to spread to
the outer edge is given by $t_T \approx N \tturb \approx \langle
\tturb \rangle [\jdisk / (\Delta j)_T]^2$, where $\jdisk$ is the
angular momentum at the outer disk edge $\rdisk$.  If we take $\rdisk$
= 100 AU, with corresponding angular momentum $\jdisk$, the timescale 
$t_T \sim 8$ Myr. However, the time required for the first planet to 
reach the outer edge can be much shorter ($\sim$ 0.1 Myr). Edge effects 
start to be important at an intermediate time scale, i.e., about 1 Myr 
(from the geometric mean). 

These time scales thus frame the problem: Since the Type I migration
time scale is roughly comparable to --- but shorter than --- the
outward diffusion time scale, the population of planetary cores is
expected to be highly depleted, even though diffusion acts to save
some fraction of them. In addition, the diffusion time scale is
comparable to expected disk lifetimes, so that the outer disk edge
will have an important impact on the results.  Keep in mind that the
time scales quoted here depend on the starting radius, the outer disk
radius, and the size of the torque parameters, so that a range of
values will be applicable to the actual population of planet-forming
disks. In particular, if the starting radius is larger (than 10 AU as
assumed above), the outward diffusion time scale will decrease and the
inward Type I migration time will increase.

\section{FOKKER-PLANCK TREATMENT: ANALYTIC RESULTS} 

After formulating this planet migration problem in terms of a
Fokker-Planck equation (Section 3.1), we explore analytic solutions.
If we consider either Type I migration torques (Section 3.2) or the
turbulent diffusion (Section 3.3) acting alone, the resulting dynamics
can be solved exactly.  We also construct a self-similar model of the
diffusion process (Section 3.4) that applies in the absence of an
outer edge to the disk.  These cases -- exact solutions to partial
versions of the problem -- provide us with an understanding of the
relevant physical mechanisms. On the other hand, they do not provide
reliable estimates for the planetary survival probabilities; these
quantities are thus determined numerically in Section 4.

\subsection{Formulation} 

Let $P(j,t)$ denote the distribution of an ensemble of planets as a
function of time.  The general form of the Fokker-Planck equation 
(e.g., Risken 1984) for this problem is given by 
\be
{\partial P \over \partial t} - 
{\partial \over \partial j} \left[ T_1(j) P \right] = 
{\partial^2 \over \partial j^2} \left[ D(j) P \right] \, , 
\ee
where $T_1(j)$ is the Type I migration torque and $D(j)$ is the
appropriate diffusion parameter due to turbulent fluctuations.  In
this problem (see also JGM), the diffusion constant is defined to be
$D \equiv (\Delta J)_T^2/\tturb$, where the fluctuation amplitude
$(\Delta J)_T$ and the time scale $\tturb$ over which the turbulent
perturbations are independent are specified in Section 2.2 (see
equations [\ref{tturbdef} -- \ref{ampturb}]).  Notice also that the
minus sign in the Type I term is included so that $T_1$ is the
magnitude of the torque.

Next we want to formulate the problem in terms of simplified 
quantities. We define a dimensionless angular momentum variable
\be
x \equiv j/j_1 \, , 
\ee
where $j_1$ is the angular momentum at a convenient reference
location; for the sake of definiteness we take $j_1 = j (r_1)$, where
$r_1$ = 1 AU.  For most cases of interest, both the torque $T_1 (j)$
and the diffusion ``constant'' $D(j)$ are functions of angular
momentum. If we specialize to the case where the disk surface density
and temperature profile are power-laws in radius (equation
[\ref{powerlaw}]), we can write $T_1 \propto j^{-a}$ and $D \propto
j^b$, where $a$ = 2 and $b$ = 1 for standard disk parameters. In
general, the indices are given by
\be
a = 2(p-q) \qquad {\rm and} \qquad b = 7 - 4p \, , 
\ee
where $p$ and $q$ are the power-law indices of the disk surface
density and temperature profiles, respectively. Next we define a
reduced Type I torque constant $\gamma$ and a reduced diffusion 
constant $\diffcon$,  
\be
\gamma \equiv {T_1 (j_1) \over j_1} \qquad {\rm and} \qquad 
\diffcon \equiv {D(j_1) \over j_1^2} \, . 
\label{nondim} 
\ee
The general form of the Fokker-Planck equation thus becomes 
\be
{\partial P \over \partial t} = \gamma 
{\partial \over \partial x} \left[ x^{-a} P \right] + 
\diffcon {\partial^2 \over \partial x^2} \left[ x^b P \right] \, .
\label{fpgeneral} 
\ee

Note that both of the constants $\gamma$ and $\diffcon$ are rates,
i.e., they have units of (time)$^{-1}$.  The Type I migration
parameter $\gamma$ takes the form
\be
\gamma = \left[ \pi f_1 \left( {m_P \over M_\ast} \right) 
\left( {r \over H} \right)^2 {G \Sigma r \over \sqrt{G M_\ast r}} 
\right]_{1{\rm AU}}  \, ,  
\label{gammadef} 
\ee
where the subscript specifies that all quantities are evaluated at $r$
= 1 AU. For typical values of the surface density at 1 AU, $\Sigma_1$
= 4500 g/cm$^2$, the scale height $(H/r)_1$ = 0.1, and for $m_P =
\sqrt{10} \mearth$, the parameter $\gamma \approx 10^{-5}$ yr$^{-1}$ =
10 Myr$^{-1}$. The reduced diffusion parameter $\diffcon$ has the form
\be
\diffcon = \left[ f_\alpha f_T^2 (2 \pi)^3 \left( 
{\Sigma r^2 \over M_\ast} \right)^2 \Omega \right]_{1{\rm AU}} \, . 
\label{betadef} 
\ee
For the same disk parameters quoted above, the value of the diffusion
parameter $\diffcon \approx 10^{-6}$ yr$^{-1}$ = 1 Myr$^{-1}$.  
The corresponding time scales are thus given by $1/\gamma \sim$ 0.1
Myr and $1/\diffcon \sim$ 1 Myr. Notice that the ratio of the Type I
torque parameter to the diffusion parameter is the ratio of time
scales given by equation (\ref{timeratio}) so that 
$\diffcon/\gamma = \qm$ (see equation [\ref{qdef}]).

In this treatment, the Fokker-Planck equation (\ref{fpgeneral}) does
not contain a source term. Although a given circumstellar disk may
produce multiple planetary cores, it will not produce a statistically
significant distribution of cores. The distribution function $P(t,x)$
thus represents the output from a large ensemble of planet-forming
disks, all with the same properties. Since this treatment does not
include planet-planet interactions, multiple cores in a particular
disk will act (statistically) as part of this same ensemble.

In addition to specifying the disk properties, we must also specify
the initial conditions, which is determined by the initial
distribution of planets $P(t=0,x)$. For most of this work, we take the
initial distribution to be a narrow gaussian centered on a given value
of angular momentum $x_0$.  Realistic disks will produce planetary
cores at a range of radial locations and hence a range of $x_0$. By
taking the initial conditions to be a narrow gaussian, we are thus
studying the effects of one starting point at a time.

Finally, we must specify the boundary conditions. At the outer edge of
the disk, corresponding to the maximum value $\xmax$ of dimensionless
angular momentum, we assume that the probability current $S(x,t)$ must
vanish. This condition is equivalent to that of requiring ``zero flux'' 
through the outer boundary and can be written in the form
\be
S(\xmax,t) = - \left[ \gamma x^{-a} P + \diffcon 
{\partial \over \partial x} \left( x^b P \right) 
\right]_{\xmax} = 0 \, , 
\label{zeroflux}
\ee
where the current $S(x,t)$ is determined by the right hand side of the
Fokker-Planck equation (\ref{fpgeneral}). In physical terms, this
boundary condition assumes that no planets can migrate beyond the
regions where disk material resides, and that no planets enter the
disk from large radii (see JGM for further discussion of this
issue). At the inner boundary, we use the ansatz $P$ = 
{\sl constant}, the form appropriate for an ``absorbing wall'' 
(Risken 1984). This inner boundary condition thus assumes that
planetary cores are accreted once they reach the star.

The survival probability for planets, and other results of interest,
depend on this choice for the outer boundary condition, as well as the
location of the outer boundary. For the special case of self-similar
solutions (Sections 3.4 and 4.2), we take the limit $\xmax \to \infty$
and apply the boundary condition (\ref{zeroflux}) there.  For disks
with a finite radial extent, our boundary condition at the outer disk
edge represents a ``reflecting wall'' (Risken 1984).  Note that it
remains possible for planetary cores near the outer edge to be
scattered outside the disk by turbulence.  Once outside the disk
material, these planets would become stranded.  If the disk edge moves
out with time, due to viscous spreading, stranded planets could be
pushed further outwards.  Although planetary cores could be ``saved''
in this manner, they would be unlikely to form giant planets because
of the lack of gas and the long orbit times (both of which inhibit
giant planet formation).  In addition, the relative amplitude is small
at the outer edge, $([\Delta J]/J)_T \approx 10^{-2}$ (see equation
[\ref{ampturb}]), so that such events could be rare.  However, if Type
I migration can reverse its direction and move planets outward (as
suggested by Paardekooper \& Mellema 2006), then this mechanism could
be important.  To include this effect in the calculations, one would
use an absorbing boundary condition at the outer edge (or a partial
barrier); this choice would allow more planetary cores to survive, but
would result in the formation of fewer giant planets.  For the
relatively short timescales of interest here ($\sim10$ Myr), the
choice of boundary condition produces modest differences; however, the
predicted survival probabilities would be affected over sufficiently
long times.

As written, equation (\ref{fpgeneral}) contains two parameters
($\gamma, \diffcon$) that set the strength of the torques and two
indices $(a,b)$ that determine their radial dependence.  Although this
formulation thus results in a four-dimensional parameter space, its
size can be reduced. First, we note that the indices $(a,b)$ have
relatively limited ranges, and that the effects of turbulence always
grow with radius compared to Type I torques. As a result, we fix
the indices to their ``standard'' values $(a,b) = (1,2)$ for much of
our exploration. For given values of the indices, one of the remaining
variables can be scaled out of the problem by changing the definition
of time. For example, let $t \to \gamma t$, and $\diffcon \to
\diffcon/\gamma = \qm$. In this case, time is measured in units of the
Type I migration time (typically several Myr) and $\qm = \diffcon / 
\gamma$ defines the level of turbulence relative to the Type I torque
strength (at 1 AU). In this reduced view, the Fokker-Planck equation
has a one parameter family of solutions, and that parameter can be 
taken to be $\qm$ as defined by equation (\ref{qdef}).  In the limit
$\qm \to 0$, Type I torques dominate the migration process, and fully
analytic solutions can be obtained (see Section 3.2). In the opposite
limit $\qm \to \infty$, turbulent torques dominate, and analytic
solutions can once again be constructed (Section 3.3).

\subsection{Solutions with Only Inward Migration} 

This section considers the limit $\qm \to 0$ where Type I torques
dominate.  In terms of the reduced quantities defined above, the
Fokker-Planck equation in the absence of diffusion has the form
\be
{\partial P \over \partial t} = 
\gamma \, {\partial \over \partial x} 
\left[ {P \over x^a} \right] \, .
\ee
General solutions of this equation can be found by making the
following transformation of both the angular momentum variable 
$x$ and the function $P$ itself:
\be
z \equiv {x^{a + 1} \over a + 1} \qquad {\rm and} \qquad
f(z,t) = x^{-a} P \left[ x(z), t \right] \, . 
\label{ztox} 
\ee
With this change of variables, the equation of motion becomes 
\be
{\partial f \over \partial t} = \gamma \, 
{\partial f \over \partial z} \, , 
\ee
which has solutions of the form 
\be
f = f (z + \gamma t) \qquad {\rm and} \qquad 
P = x^{a} f(z + \gamma t) \, , 
\ee
where $z$ is related to $x$ through equation (\ref{ztox}).  The form
of the function $f$ is specified by the initial condition, so that
\be
f(z) = x^{-a} P(x,0) \, . 
\ee

To illustrate this type of solution, we consider the case where the
initial distribution of angular momentum has a gaussian form, i.e.,
\be
P(x,0) = {1 \over \sigma \sqrt{\pi} } \, 
\exp \left[ - {(x - x_0)^2 \over \sigma^2} \right] \, , 
\ee
where $x_0$ is the angular momentum at the peak of the initial
distribution.  Note that the distribution is normalized over positive
angular momentum values $x$ and has width given by $\sigma$.  The
time-dependent solution thus has the form
\be
P(x,t) = {1 \over \sigma \sqrt{\pi} } \, 
{x^a \over [x^{a+1} + (a+1)\gamma t]^{a/(a+1)} } \, 
\exp \left[ - { \left\{ [x^{a+1} + (a+1)\gamma t]^{1/(a+1)} 
- x_0 \right\}^2 \over \sigma^2 } \right] \, . 
\label{ingeneral} 
\ee
The probability $p_S (t)$ of planet survival can be found by
integrating the solution given by equation (\ref{ingeneral}) over all
positive values of $x$. In the limit where the width of the initial
distribution is small compared to the peak, and the time of
observation is long, the parameters of the problem obey the ordering
\be
[ (a+1) \gamma t)]^{1/(a+1)} \gg x_0 \gg \sigma \, . 
\ee
In this limit, the survival probability can be written in the form 
\be
p_S (t) = {1 \over 2} {\rm Erfc} \left[ 
{ [ (a+1) \gamma t)]^{1/(a+1)} \over \sigma } \right] \approx 
{\sigma \over 2 \sqrt{\pi} [ (a+1) \gamma t)]^{1/(a+1)} }
\exp \left[ - { [ (a+1) \gamma t)]^{2/(a+1)} 
\over \sigma^2 } \right] \, ,
\ee
where Erfc$(x)$ is the complementary error function (AS), 
and where the second approximate equality holds in the 
asymptotic limit.  

\subsection{Solutions with Only Diffusion} 

This section considers the opposite limit where $\qm \to \infty$,
i.e., we neglect the Type I migration torques so that $\gamma$ = 0.
We make the additional restriction to the case where the diffusion
constant $D(j) \propto j$ (so that $b$ =1 ); as a result, this
treatment is not as general as that of Section 3.2.  If we redefine
the time variable so that $t \to \diffcon t$, the diffusion equation
for the probability distribution takes the form 
\be
{\partial P \over \partial t} = 
{\partial^2 \over \partial x^2} \left[ x P \right] \, . 
\ee
Note that $t$ is a dimensionless time variable, or, equivalently, 
time is measured in units of the diffusion timescale. 

If we separate the diffusion equation so that $P(x,t) = G(t) F(x)$, 
the temporal solutions take the form 
\be
G(t) = \exp \left[ - \lambda t \right] \, , 
\ee
where $\lambda$ is the separation constant, and the remaining
differential equation for $F(x)$ becomes
\be
x {d^2 F \over dx^2} + 2 {dF \over dx} + \lambda F  = 0 \, .
\label{space} 
\ee
After some rearrangement, the solution to equation (\ref{space}) 
can be written in the form 
\be
F(x) = {1 \over \sqrt{\lambda x} } J_1 \left( 2 \sqrt{\lambda x} \right) \, , 
\ee
where $J_1 (x)$ is the Bessel Function of the first kind of order one
(Abramowitz \& Stegun 1970; hereafter AS).  This solution is chosen to
be finite at the origin $x$ = 0. To apply the outer boundary
condition, we require that the flux at the outer edge of the disk
vanish. This location corresponds to a maximum value $\xmax$ of the
dimensionless angular momentum. After defining $\xi \equiv 2$ 
$\sqrt{\lambda x}$, the outer boundary condition (see equation 
[\ref{zeroflux}]) takes the form
\be
{d \over d x} \left[ x F(x) \right] = 0 \qquad \Rightarrow \qquad 
{d \over d\xi} \left[ \xi J_1 (\xi) \right] = 0 = \xi J_0 (\xi) \, , 
\ee
where we have used the properties of Bessel functions (AS) to obtain
the final equality. The separation constant must be chosen so that the
outer boundary occurs at a zero of the zeroth order Bessel function 
$J_0$. If we denote the zeroes of $J_0$ by $\xi_\nu$, the separation
constants $\lambda_\nu$ are given by
\be
\lambda_\nu = {\xi_\nu^2 / 4 \xmax} \, . 
\label{lambda}
\ee
The general solution thus takes the form 
\be
P(x,t) = \sum_{\nu =1}^\infty A_\nu \exp \left[ - \lambda_\nu t \right] 
{1 \over \sqrt{\lambda_\nu x} } J_1 \left( 2 \sqrt{\lambda_\nu x} \right) \, , 
\label{fullp} 
\ee
where the $\lambda_\nu$ are given by equation (\ref{lambda}). 
Suppose we are given an initial distribution $f(x)$ at $t$ = 0, i.e., 
\be
P(x,t=0) = \sum_{\nu =1}^\infty A_\nu {1 \over \sqrt{\lambda_\nu x} } 
J_1 \left( 2 \sqrt{\lambda_\nu x} \right) = f(x) \, . 
\ee
Next we multiply both sides of the equation by 
$\sqrt{x} \, J_1 (2 \sqrt{\lambda_\mu x})$ and then integrate:
\be
\sum_{\nu =1}^\infty {A_\nu \over \sqrt{\lambda_\nu} } 
\int_0^\xmax dx J_1 \left( 2 \sqrt{\lambda_\nu x} \right) 
J_1 \left( 2 \sqrt{\lambda_\mu x} \right) = 
\sum_{\nu =1}^\infty {A_\nu \over \sqrt{\lambda_\nu} } I_\nu = 
\int_0^\xmax \, dx \, \sqrt{x} \, f(x) \, 
J_1 \left( 2 \sqrt{\lambda_\mu x} \right) \, . 
\ee
The integrals $I_\nu$ in the sum can then be rewritten by changing 
variables to $u^2 = x/\xmax$, so they take the form 
\be
I_\nu = 2 \xmax \int_0^1 \, u \, du \, J_1 (\xi_\nu u) J_1 (\xi_\mu u) \, , 
\ee
where the $\xi_k$ are zeroes of the $J_0$ functions. After applying the 
recursion relations for Bessel functions and integrating by parts, we find 
\be
I_\nu = 2 \xmax {\xi_\nu \over \xi_\mu} \int_0^1 \, u \, du \, 
J_0 (\xi_\nu u) J_0 (\xi_\mu u) \, = \xmax 
\left[ J_1 (\xi_\nu) \right]^2 \delta_{\nu \mu} \, . 
\ee
The coefficients $A_\nu$ can now be evaluated: 
\be
A_\nu = {\sqrt{\lambda_\nu} \over \left[ J_1 (\xi_\nu) \right]^2} 
\, {1 \over \xmax} \int_0^\xmax \, dx \, \sqrt{x} \, f(x) \, 
J_1 \left( 2 \sqrt{\lambda_\nu x} \right) \, . 
\ee

As one example, we consider the case in which all of the planets start
at the same radius, or angular momentum, so that the starting
distribution $f(x) = \delta (x - x_0)$, and the $A_\nu$ take the form
\be
A_\nu = {\sqrt{\lambda_\nu x_0} \over \xmax \left[ J_1 (\xi_\nu) \right]^2} 
J_1 \left( 2 \sqrt{\lambda_\nu x_0} \right) = 
{\xi_\nu \sqrt{x_0/\xmax} \over 2 \xmax \left[ J_1 (\xi_\nu) \right]^2} 
J_1 \left( \xi_\nu \sqrt{x_0/\xmax} \right) \, . 
\label{adelta} 
\ee

Next, we can find the total survival probability by starting with the
full time dependent solution of equation (\ref{fullp}) and integrating
over all angular momentum values,
\be
p_S (t) = \int_0^\xmax P(x,t) dx = \sum_{\nu =1}^\infty A_\nu 
\exp \left[ - \lambda_\nu t \right] \int_0^\xmax \, 
{dx \over \sqrt{\lambda_\nu x} } J_1 \left( 2 \sqrt{\lambda_\nu x} \right) 
= \sum_{\nu =1}^\infty {4 \xmax A_\nu \over \xi_\nu^2} 
\exp \left[ - \lambda_\nu t \right] \, . 
\ee
For example, for the particular case in which the starting angular
momentum distribution is a delta function, so that the coefficients
$A_\nu$ are given by equation (\ref{adelta}), the probability takes 
the form 
\be
p_S (t) = \sum_{\nu =1}^\infty 
{2 u_0 J_1 (\xi_\nu u_0) \over \xi_\nu \left[ J_1 (\xi_\nu) \right]^2} 
\exp \left[ - \lambda_\nu t \right] \, , 
\label{pvst} 
\ee
where $u_0 \equiv (x_0 / \xmax)^{1/2}$. 

In this problem, the flux at the outer boundary vanishes, 
and the flux into the origin is given by 
$$
{\cal F}_0 = - {\partial \over \partial x} \left( x P \right) 
\Bigg|_{x=0} =
- \sum_{\nu =1}^\infty A_\nu \exp \left[ - \lambda_\nu t \right] 
{1 \over \sqrt{\lambda_\nu} } {\partial \over \partial x} 
\left[ x^{1/2} J_1 \left( 2 \sqrt{\lambda_\nu x} \right) 
\right]_{x=0} \,
$$
\be
\, \qquad = - 
\sum_{\nu =1}^\infty A_\nu \exp \left[ - \lambda_\nu t \right] 
\left[ {J_1 \left( x_\nu \right) \over x_\nu} + 
{d J_1 \over dx} \right]_{x_\nu = 0} \, = -
\sum_{\nu =1}^\infty A_\nu \exp \left[ - \lambda_\nu t \right] \, , 
\ee
where $x_\nu = 2 \sqrt{\lambda_\nu x}$. For comparison, 
\be
{d p_S \over dt} = - \sum_{\nu =1}^\infty 
A_\nu {4 \xmax \lambda_\nu \over \xi_\nu^2} 
\exp \left[ - \lambda_\nu t \right] 
= - \sum_{\nu =1}^\infty 
A_\nu \exp \left[ - \lambda_\nu t \right] \, . 
\ee
Thus, $d p_S / dt = {\cal F}_0$, as expected. 

At late times, only the leading term survives in the series that
describes the solutions.  As a result, the first term of equation
(\ref{fullp}) determines the probability distribution in the long time
limit. As a result, the distribution of locations for surviving
planetary cores is given by the first order Bessel function of the
first kind. Similarly, the total survival probability is given by the
first term in equation (\ref{pvst}). The first three zeroes (of the
zeroth Bessel function of the first kind) are $\xi_1 \approx$
2.40482, $\xi_2 \approx$ 5.52007, and $\xi_3 \approx$ 8.65372 (AS); 
if we take the outer boundary to be $\xmax$ = 10 (corresponding 
to an outer disk radius of 100 AU), the first three eigenvalues (see
equation [\ref{lambda}]) are approximately $\lambda_1 \approx$ 0.145,
$\lambda_2 \approx$ 0.762, and $\lambda_3 \approx$ 1.87.  After one
diffusion time scale (roughly 1 Myr), the first term is about twice as
large as the second. After 10 diffusion times (about 10 Myr), the
first term is almost 500 times larger. In the (expected) case in which
the planetary core population is severely depleted, the distributions
are thus determined primarily by the leading order terms. We exploit
this property of the solutions in Section 4.3, which determines the 
lowest order eigenfunctions and eigenvalues for the full problem,
including Type I migration. 

\subsection{Self-Similar Solutions} 

In the absence of an outer disk edge, self-similar solutions to the
Fokker-Planck equation exist (JGM).  Although we expect the disk
radius to be finite, with typical radii $\rdisk \sim$ 30 -- 100 AU, we
can use self-similar solutions as an analytic model of the dynamics to
gain further insight into the problem. One should keep in mind,
however, that these solutions overestimate the probability of
planetary survival.

In the limit of long times, the surviving planets tend to reside in
the outer disk where inward migration due to Type I torques is
relatively unimportant compared with diffusion. As shown previously,
self-similar solutions exist in this regime when the Type I torques
vanish (JGM).  However, we can include an inward torque term, and
still retain self-similarity, provided that we use an averaged torque
so that the Fokker-Planck equation takes the form 
\be
{\partial P \over \partial t} = \gambar {\partial P \over \partial x} 
+ \diffcon {\partial^2 \over \partial x^2} \left( x P \right) \, , 
\label{fpselfsim} 
\ee
where $\gambar$ is now an appropriate average over the disk (to remove
the additional $x$-dependence in the torque term).  Since Type I
torques remove planets from the inner disk on a short time scale
(compared with the disk lifetime), the effective value $\gambar$
should be representative of the outer disk; as a reference point, we
expect $\gamma \approx \diffcon$ (0.1 $\diffcon$) at $r \approx$ 10 AU
(100 AU).  Although this model equation is simpler than the full
problem, it retains the crucial feature that the relative importance 
of diffusion (compared with Type I migration) increases outwards. In 
addition, an analytic solution can be found (see below) and the
optimization calculation (see Section 4.2) can be done explicitly. 

This version of the Fokker-Planck equation (\ref{fpselfsim}) 
has the solution
\be
P(x,t) = A (\diffcon t + \sigma)^{-(2 + \gambar/\diffcon)} 
\exp[-x/(\diffcon t + \sigma)] \, , 
\label{selfsimsol} 
\ee
where $\sigma$ is a constant that is determined by the initial width
of the distribution, and where $A$ is a normalization constant.  For
standard normalization, the expectation value of the initial state is
given by $\langle x_0 \rangle = \sigma$. For planets starting near $r$
= 10 AU, we expect $\sigma \sim x_0 \sim \sqrt{10} \sim 3$.  As
written, the solution extends to spatial infinity ($x \to \infty$),
where the distribution function obeys a zero-flux outer boundary
condition (equation [\ref{zeroflux}]). Keep in mind that the solution
given in equation (\ref{selfsimsol}) is the simplest member of a
sequence of self-similar solutions.

In this model, the probability of a planet remaining in the disk is
given by integrating the above solution over all values of $x$, 
\be
p_S(t) = \int_0^\infty dx P(x,t) = {A \over 
(\diffcon t + \sigma)^{1 + \gambar/\diffcon} } \, \approx 
\left( 1 + \diffcon t /\sigma \right)^{-(1 + \gambar/\diffcon)} \, , 
\label{selfsimint} 
\ee
where $A$ is a normalization constant, and we have normalized the
solution in the final equality so that the total probability is unity
at $t$ = 0. This result provides an exact solution to the simplified
problem posed by equation (\ref{fpselfsim}), but is only an
approximation to the original physical problem (with spatially varying
Type I torques) because the true solution is expected to approach the
self-similar form of equation (\ref{selfsimsol}) only at late times. 
As a result, the normalization (defined here at $t$ = 0) can be
different.

In the limit $\gambar \to 0$, the survival fraction approaches the
form $p_S \propto 1/t$ (see JGM).  When Type I torques are included,
the power-law steepens and hence fewer planets survive. Notice that
this solution represents an upper bound on the true survival fraction
for two reasons: The inclusion of the outer boundary (at the disk
edge) enforces exponential decay in the long term (see Sections 3.3,
4.3, and Figures \ref{fig:ptime} and \ref{fig:lvg}).  In addition, the
Type I torques are approximated here with no spatial dependence, and
hence take on the value appropriate in the outer disk; including the
spatial dependence will increase their efficacy and hasten the removal
of planets from the inner disk.

Nonetheless, we can obtain a working estimate for the survival
fraction. For typical values $\diffcon$ = 1 Myr$^{-1}$, 
$\gambar/\diffcon$ = 0.3, distribution width $\sigma$ = 3 = 
$\langle x_0 \rangle$, and time $t$ = 10 Myr, we find $p_S \approx$
0.15. For this time scale, we thus find that planetary survival is
only moderately rare, at the level of ten percent, roughly consistent
with the numerical calculations of the previous section. This estimate
is somewhat higher, however, primarily due to the absence of the outer
boundary.

\section{FOKKER-PLANCK TREATMENT: NUMERICAL RESULTS} 

We can directly solve the Fokker-Planck equation using standard
numerical methods; here we use a fully implicit method (e.g., see
Press et al. 1990).  The boundary conditions play an important role in
determining the fraction of surviving planets as a function of
time. For the calculations of this section, we adopt a standard set of
boundary conditions and initial conditions in order to determine how
the planet survival fractions depend on time and on the torque
parameters $\gamma$ and $\diffcon$.  The inner boundary is fixed at $x
= x_\ast = 0.1$ ($r$ = 0.01 AU) and the outer boundary is fixed at $x
= \xmax = 10$ ($r$ = $\rdisk$ = 100 AU). The distribution function $P$
is chosen to have a constant value at the inner boundary; note that a
constant value of $P$ allows for nonzero flux through the inner
boundary.  At the outer boundary, we use the zero-flux condition,
which in this formulation is given by equation (\ref{zeroflux}). The
initial condition is chosen to be a narrow gaussian distribution
centered on $x = x_p = \sqrt{r_0}$, corresponding to radius $r_0$.  We
use $r_0$ = 10 AU as a benchmark value, but explore varying values.

Through numerical experimentation, we find that the width of the
initial gaussian has relatively little effect, provided that it is
much narrower than the disk size.  The location of the peak determines
two important time scales for the evolution of the probability
distribution: (1) the time required for the Type I torques to move
planets from the peak location inward to the star, and (2) the time
required for diffusion to spread the distribution to the outer disk
edge, where the outer boundary affects the dynamics. 

\subsection{Basic Numerical Results} 

\begin{figure} 
\figurenum{1}
{\centerline{\epsscale{0.90} \plotone{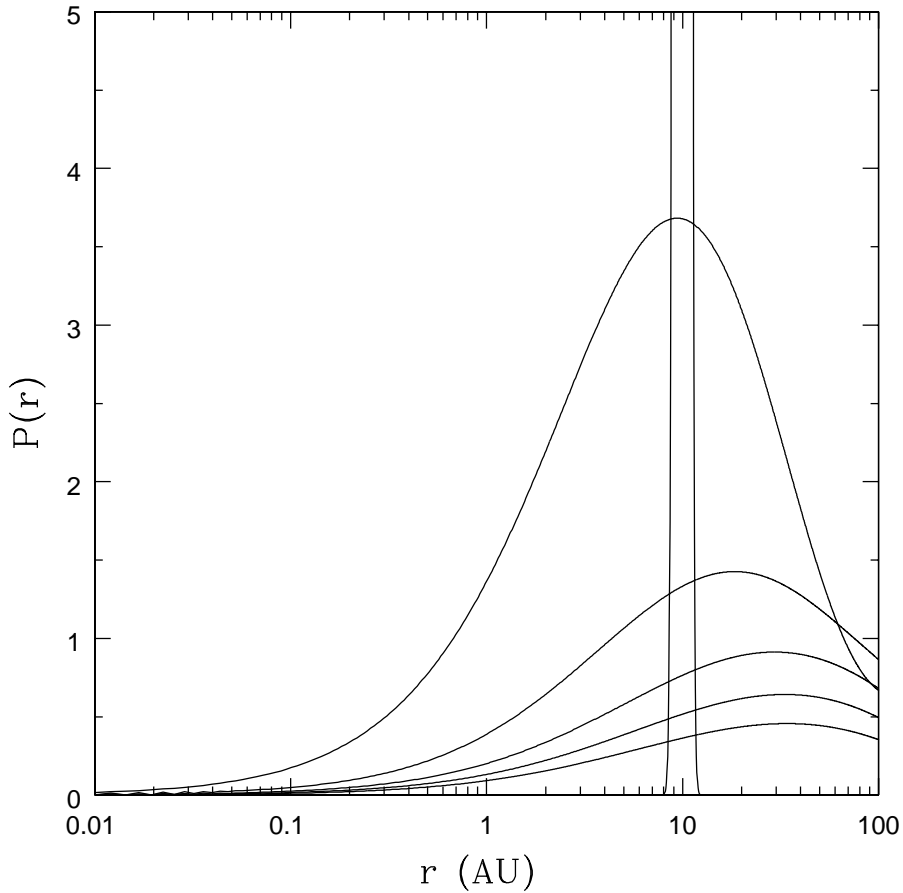} } } 
\figcaption{Distributions of radial locations of migrating planets
from numerical solution to the Fokker-Planck equation. The solution 
$P$ is the distribution function for the dimensionless angular
momentum $x$, so that $P = dp/dx$, but the result is plotted versus
radius $r \propto x^2$. The solutions are shown at six sampling times:
0, 1, 2, 3, 4, and 5 Myr, from top to bottom in the figure. The torque
parameters are chosen to be $\gamma$ = 10 Myr$^{-1}$ and $\diffcon$ =
1 Myr$^{-1}$. Recall that $\gamma \propto m_P \Sigma/H^2$ (see
equation [\ref{gammadef}]) and that $\diffcon \propto \Sigma^2$ 
(see equation [\ref{betadef}]). } 
\label{fig:pdist} 
\end{figure}

The evolution of the probability distribution is illustrated in Figure
\ref{fig:pdist}.  The torque parameters are chosen to be near the
center of the range of expected values with $1/\gamma$ = 0.1 Myr and
$1/\diffcon$ = 1 Myr.  The figure shows the distribution $P(r,t)$ as a
function of radius $r$. Note that the calculations are done in terms
of dimensionless angular momentum $x$, so that the function $P$
represents the probability density in $x$, i.e., $P = dp/dx$. In the
figures of this paper, however, we plot the function $P$ versus radius
$r = x^2$ AU (because we have better intuition for the meaning of
radial locations in these disks).  As expected, the distribution
spreads out with time, and its area decreases as planets are lost
through accretion onto the central star. The peak of the distribution
actually moves outwards with time, even though Type I migration acts
to move planets inward. Here, at the relatively late times shown, any
planets that diffuse into the inner regions of the disk are quickly
swept into the star, and thus do not contribute to the distribution at
small radii.  Notice that the edge of the probability distribution
reaches the outer boundary in only about 1 Myr, so the effect of the
outer disk edge plays an important role in determining planet
survivability on this time scale (and longer).

\begin{figure} 
\figurenum{2}
{\centerline{\epsscale{0.90} \plotone{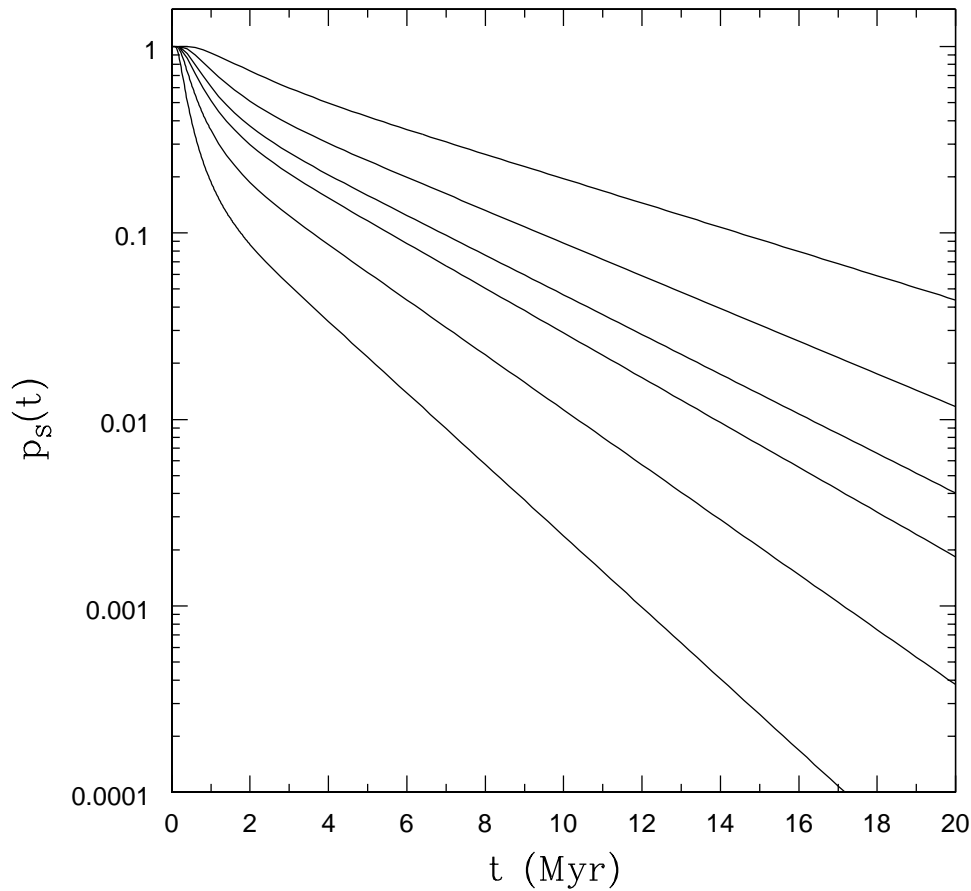} } } 
\figcaption{Time evolution of the fraction of surviving planets from
numerical solution to the Fokker-Planck equation.  The curves show the
results using varying values of the parameter $\gamma$ that sets the
rate of Type I migration relative to the level of turbulence.  The
values are $\gamma$ = 0, 1, 3, 5, 10, and 20 from top to bottom
(in units of Myr$^{-1}$). The Type I migration parameter $\gamma$ 
scales linearly with both the planetary core mass $m_P$ and with 
the disk surface density $\Sigma$. }
\label{fig:ptime} 
\end{figure}

For a given distribution $P(x,t)$ at a specific time, the fraction 
of surviving planets $p_S(t)$ is given by the integral 
\be
p_S (t) = \int_{x_\ast}^{\xmax} P(x,t) dx \, . 
\ee
Figure \ref{fig:ptime} shows the total probability of planet survival
as a function of time for varying values of the Type I migration
torques and fixed amplitude of the turbulent torques (with $\diffcon$
= 1 Myr$^{-1}$). Figure \ref{fig:ptime} is presented as a log-linear
plot, so that exponential decay corresponds to straight lines in the
diagram. Note that all of the curves become straight lines
asymptotically with time, so that the decay rate is in fact
well-defined. 
 
The solutions depicted in Figure \ref{fig:ptime} provide estimates for
the survival probability.  In the absence of diffusion, the angular
momentum of migrating planets decreases according to $x(t) = x_0 (1 -
3 \gamma t/x_0^3)^{1/3}$.  Using a typical Type I migration rate
($\gamma^{-1}$ = 0.1 Myr) and the initial conditions of the numerical
simulations (where $x_0 \approx x_p = \sqrt{10}$), the angular
momentum reaches zero (planets are accreted) in time $t_{acc} =
x_0^3/(3 \gamma) = \sqrt{10}/3 \approx 1.1$ Myr. For comparison,
when turbulent fluctuations are included at the ``standard'' level (so
that $\diffcon^{-1}$ = 1 Myr), the survival fraction is $p_S \approx
0.36$ at time $t$ = 1 Myr and $p_S \approx 0.19$ at time $t$ = 2 Myr.
The planetary survival fraction falls to $p_S = 0.10$ at time $t
\approx$ 3.6 Myr. Turbulence thus allows planets to survive several
times longer than they would otherwise.  Nonetheless, in the long time
limit, few planets survive: only about 1 percent ($p_S \approx 0.01$)
of the starting population is still present at $t$ = 10 Myr.

\begin{figure} 
\figurenum{3}
{\centerline{\epsscale{0.90} \plotone{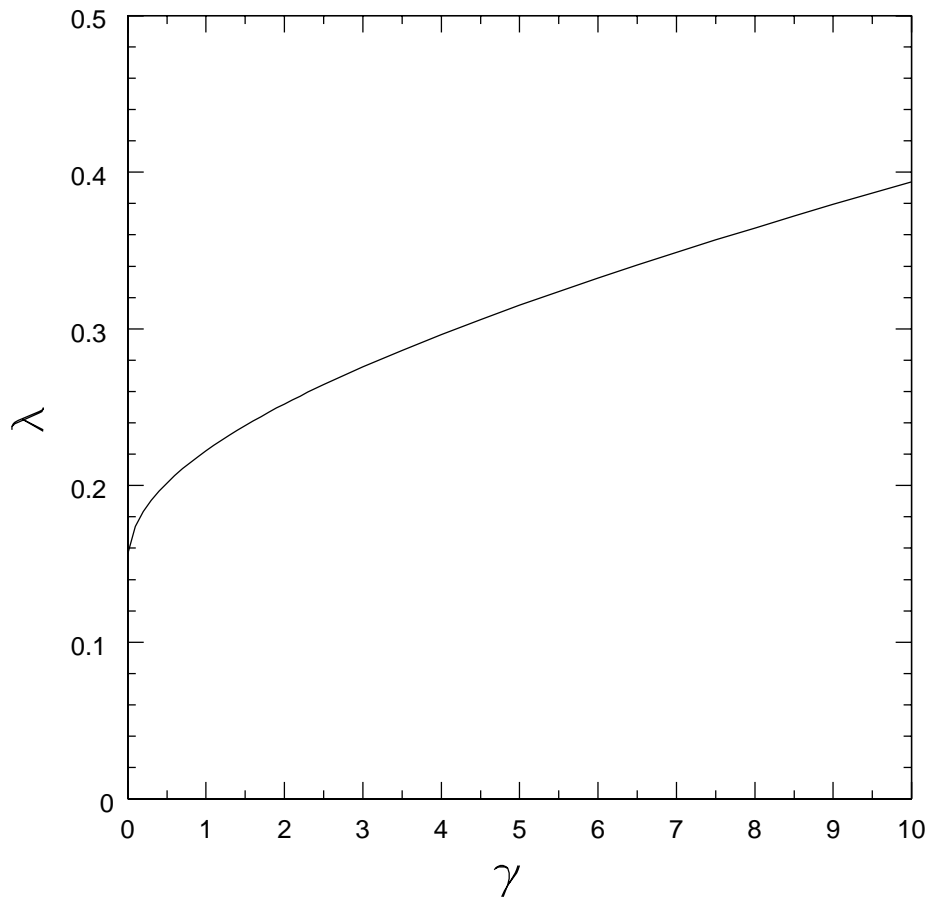} } } 
\figcaption{Exponential decay rate $\lambda$ for planet survival as 
a function of Type I migration parameter $\gamma$ for fixed diffusion
parameter (here $\diffcon$ = 1), where all quantities are given in
units of Myr$^{-1}$. The decay rates are evaluated from the numerical
solutions at an evolution time of 20 Myr. }
\label{fig:lvg} 
\end{figure}

For the same disk torque parameters used to construct Figure
\ref{fig:ptime}, the decay rate $\lambda$ is shown as a function of
the parameter $\gamma$ in Figure \ref{fig:lvg}.  For the sake of
definiteness, the decay rates $\lambda = d \ln p_S / d \ln t$ are
evaluated at time $t$ = 20 Myr. Figure \ref{fig:ptime} shows that
little curvature remains in the survival fractions at times of 20 Myr,
so that the decay rates have nearly reached their asymptotic values.
One should keep in mind, however, that some longer term evolution is
possible.  In the limit $\gamma$ = 0, the decay rate approaches the
value $\lambda \approx 0.16$, in agreement with the leading order
result $\lambda_1 \approx 0.15$ derived in Section 3.3 using roots 
of the Bessel function.

\begin{figure} 
\figurenum{4}
{\centerline{\epsscale{0.90} \plotone{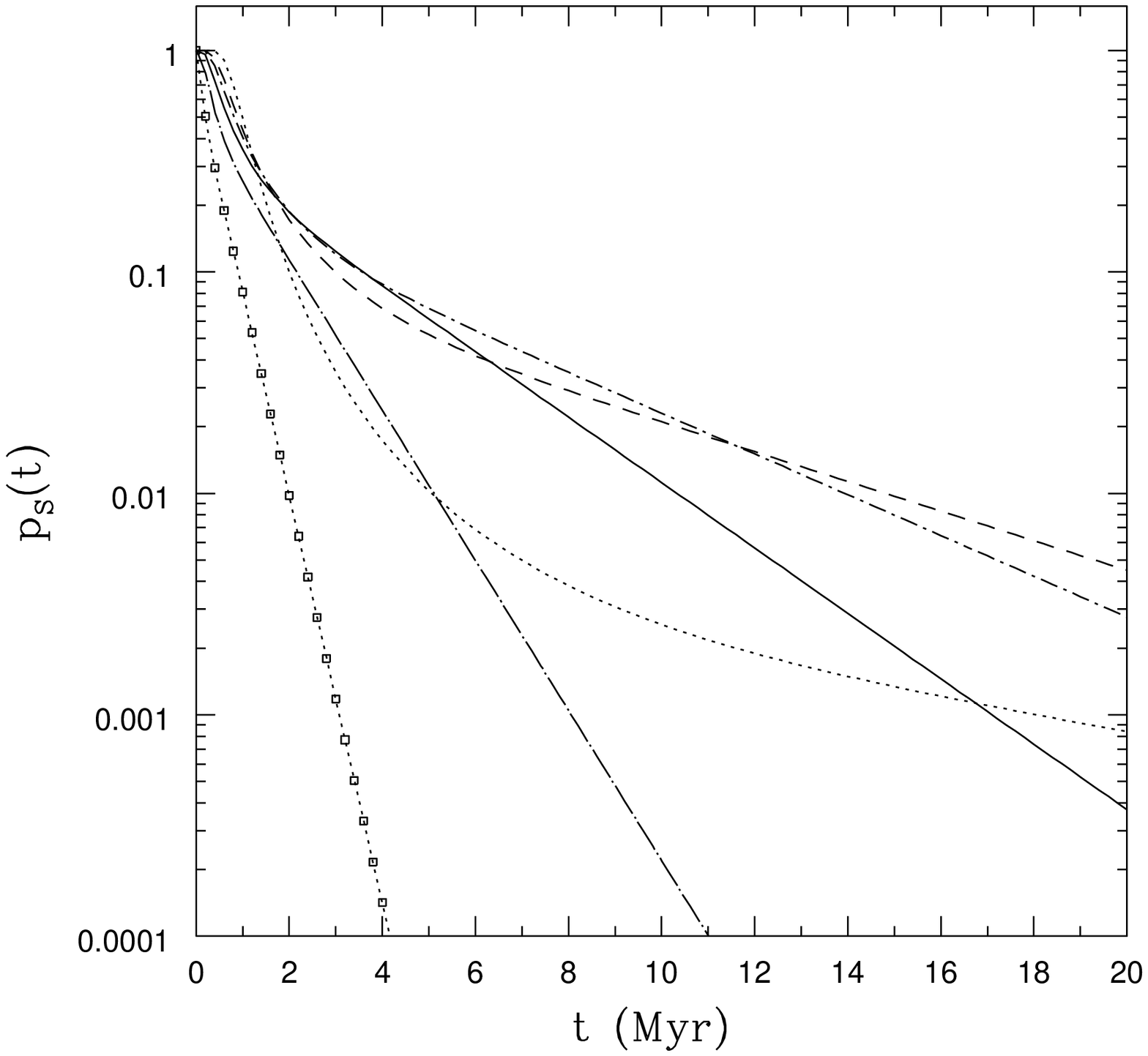} } } 
\figcaption{Time evolution of the fraction of surviving planets from
the Fokker-Planck equation using varying values of the diffusion
parameter $\diffcon$. The curves correspond to values of $\diffcon$ =
0.1 (dots), $\diffcon$ = 0.3 (dashes), $\diffcon$ = 0.5 (dot-dashes), 
$\diffcon$ = 1 (solid), $\diffcon$ = 3 (dot-long-dashes), and $\diffcon$
= 10 (dots marked by open squares). The diffusion parameters are given 
in units of Myr$^{-1}$. }
\label{fig:ptimebeta} 
\end{figure}

As another way to view these systems, we can plot the survival
fraction $p_S$ as a function of time for fixed Type I torque parameter
$\gamma$ and varying values of the diffusion parameter $\diffcon$. One
set of results is shown in Figure \ref{fig:ptimebeta} for $\gamma$ =
10 and diffusion parameter in the range $0.1 \le \diffcon \le 10$.
For relatively ``large'' diffusion parameters, corresponding to high
levels of turbulence, the survival curves show the same exponential
behavior as in Figure \ref{fig:ptime}. For $\diffcon \approx 0.3$,
however, the curves show more structure, and larger fractions of the
planetary population survive. For even smaller values of the diffusion
parameter (not shown in the Figure), turbulence has little effect, and
steady disk torques sweep (almost) the entire population of planets
into the star on the Type I migration timescale. This behavior
suggests that for a fixed value of $\gamma$, there exists an optimum
value of the diffusion parameter $\diffcon$ that maximizes the number
of surviving planets.  This optimum value depends on the time of
observation and is taken up in Section 4.2.

\begin{figure} 
\figurenum{5}
{\centerline{\epsscale{0.90} \plotone{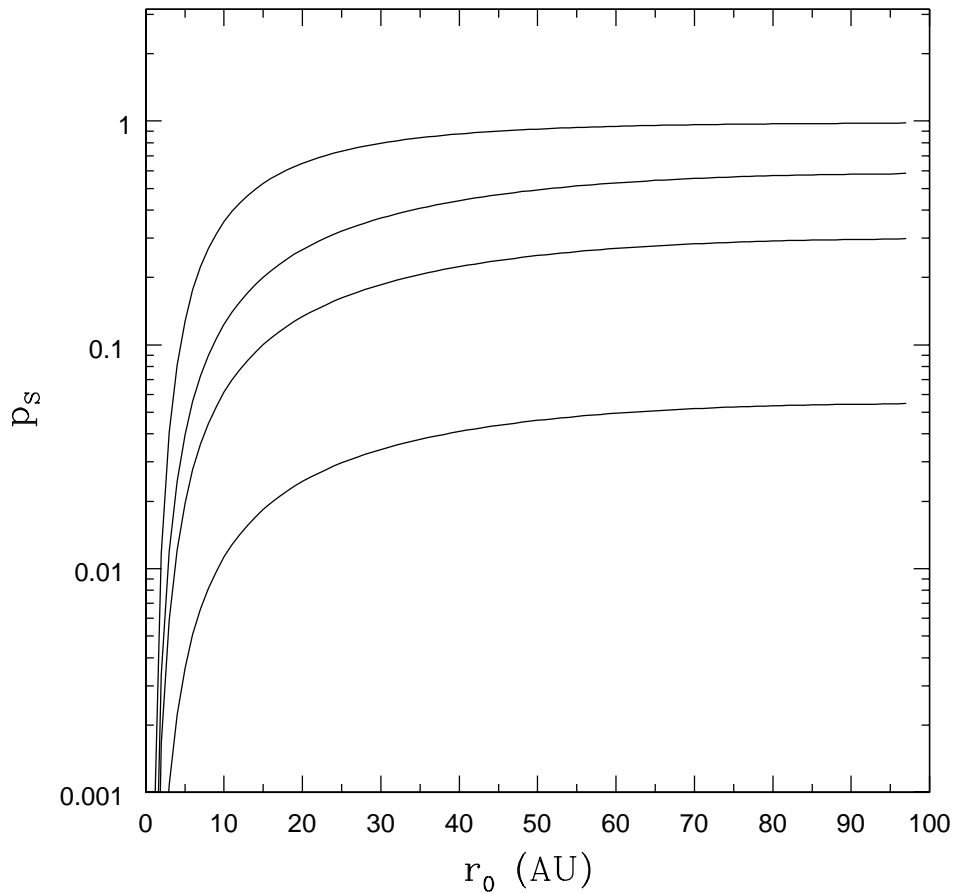} } } 
\figcaption{Planet survival fraction as a function of starting 
location. In each of these simulations, the initial distribution of
angular momenta is taken to be a narrow gaussian centered on a value
given by $x_0 = [r_0/(1 {\rm AU})]^{1/2}$. The fraction of surviving
planets is plotted as function of $r_0$ for four sampling times: $t$ 
= 1 Myr, 3 Myr, 5 Myr, and 10 Myr (from top to bottom in the figure).}  
\label{fig:rstart} 
\end{figure}

All of the results shown thus far correspond to the same initial
distribution of angular momentum, i.e., a narrow gaussian centered on
the angular momentum appropriate for a circular orbit at $r_0$ = 10
AU. We expect the planetary survival rate depend on the starting
location. As outlined in Section 2, for typical torque parameters, the
time scale for inward Type I migration and that for turbulent
diffusion are comparable for radii near 10 AU. For smaller radii, Type
I torques are dominant, and fewer planets should survive.  For larger
radii, turbulence dominates. To study this issue, we have performed a
series of simulations in which the starting location is a narrow
gaussian centered on an angular momentum value corresponding to a
range of radial locations from 1 AU to the outer disk edge. The
results are shown in Figure \ref{fig:rstart} for four sampling times
(1, 3, 5, and 10 Myr). Notice that the four curves display a sharp
corner near $r_0 \sim 10$ AU. For smaller radii, the survival fraction
$p_S$ drops precipitously. For larger radii, the fraction $p_S$ is a
fairly flat function of radius at a given sampling time.

\subsection{Optimization of Survival Probability} 

For a given Type I migration rate and a given time, there exists an
optimum value of the diffusion constant that provides the greatest
number (fraction) of surviving planets. This claim can be seen as
follows: In the limit of no diffusion $D \to 0$, the planets all
migrate inward. For times greater than the Type I migration time,
essentially all of the planets are accreted by the central star, and
the number of surviving planets approaches zero.  In the opposite
limit where the diffusion constant is large, the random walk in
angular momentum introduced by the diffusion process would lead to
crossings of the origin (where $j \to 0$, $r \to 0$, and accretion
takes place) in only a few steps.  Given the one-way barrier at the
stellar surface, the fraction of surviving planets also vanishes in
the limit of large $D$. As a result, an optimum value of the diffusion
constant can occur in the intermediate regime.

\begin{figure} 
\figurenum{6}
{\centerline{\epsscale{0.90} \plotone{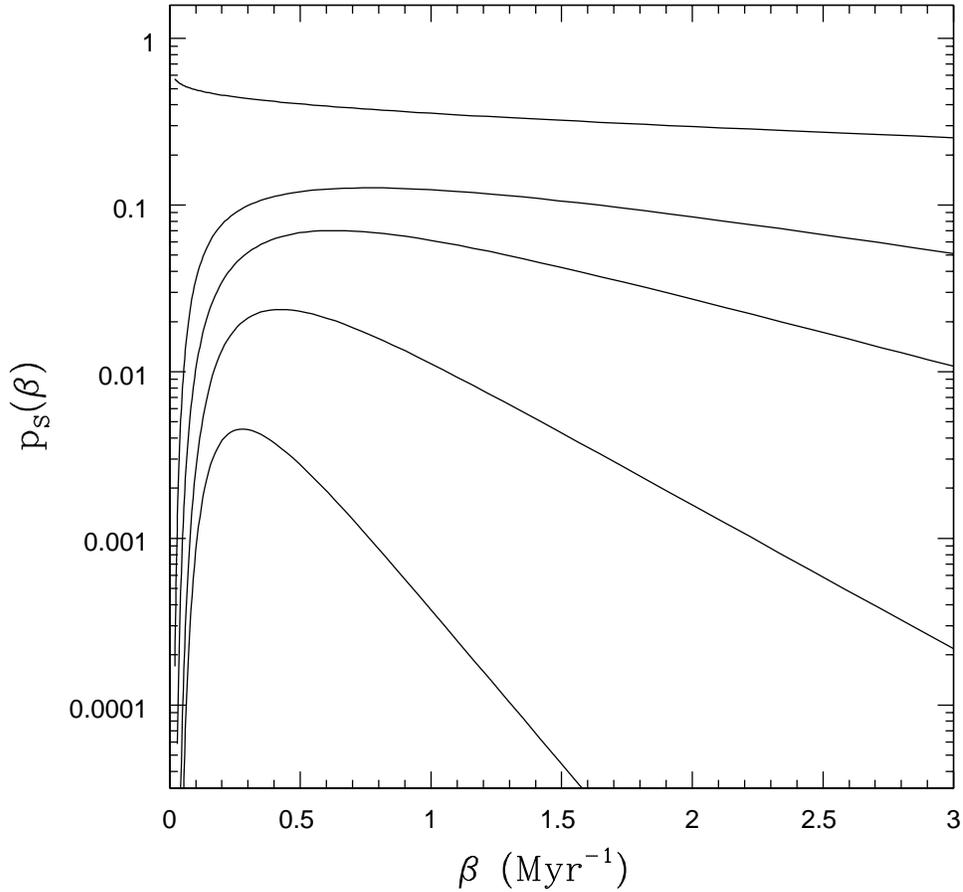} } } 
\figcaption{Total survival fraction as a function of the diffusion 
parameter $\diffcon$ (the value of the diffusion constant at $r$ = 1
AU in units of Myr$^{-1}$).  The Type I migration parameter $\gamma$
is kept constant at a value of 10 Myr$^{-1}$. The curves shown
correspond to times of 1 Myr (top), 3 Myr, 5 Myr, 10 Myr, and 20 Myr
(bottom). The initial distribution of angular momentum for this set of
simulations was a narrow gaussian centered at $x = \sqrt{10}$, i.e.,
$r$ = 10 AU. }
\label{fig:optim} 
\end{figure}

Figure \ref{fig:optim} shows the results of numerically exploring this
optimization problem. At fixed sampling times, the fraction of
surviving planets is shown as a function of the diffusion constant for
a fixed value of the Type I migration torque ($\gamma$ = 10 Myr$^{-1}$).  
At early times (the uppermost curve in the figure at $t$ = 1 Myr), the
Type I migration process has not had time to remove all of the
planets, and the result of increasing the diffusion constant is to
decrease the number of surviving bodies. At all later times shown,
however, a maximum appears in the survival fraction at intermediate
values of the diffusion constant.  Note that this maximum occurs for
values of the diffusion parameter near those expected from ``typical''
turbulent torques, although a wide range of such parameters are
possible. 

The optimum value of the diffusion parameter depends on the other
properties of the system: Here we have used the expected value of the
Type I migration parameter $\gamma$ = 10 Myr$^{-1}$ and used initial
conditions where the planetary cores are formed near $r$ = 10 AU.
However, notice that with the general form of the Fokker-Planck
equation (\ref{fpgeneral}), one can absorb the parameter $\gamma$ into
the definition of time, so that the results depend only on the ratio
$\diffcon/\gamma = \qm$ (see equation [\ref{qdef}]). Further, $\qm$
depends on the physical properties of the systems according to 
$\qm \propto \Sigma H^2 / m_P$. 

To illustrate this optimizing behavior, we consider the simplified,
self-similar version of the problem developed in Section 3.4.
Specifically, we use the self-similar solution of equation
(\ref{selfsimsol}) as a model for the dynamics.  This treatment does
not include the outer disk boundary, and hence overestimates the
survival probability. On the other hand, it provides an analytic
understanding of how the parameter space of Type I torque strength
(given here by $\gambar$) and diffusion constant $\diffcon$ contains a
local maximum in the fraction of surviving planetary cores. 

Given the normalized solution of equation (\ref{selfsimint}) for the
survival fraction $p_S$ as a function of time, we can find the
optimum value of the diffusion parameter $\diffcon$ for fixed time
$t$ and migration parameter $\gambar$.  The extremal value occurs
where $dp_S/d\diffcon$ = 0, which implies the constraint 
\be
\gambar \ln \left[ 1 + \diffcon t / \sigma \right] =  
{\diffcon t / \sigma \over 1 + \diffcon t / \sigma} 
\left( \diffcon + \gambar \right) \, . 
\label{best} 
\ee
Equation (\ref{best}) has a solution provided that the parameter
$\alpha \equiv \gambar t / \sigma \ge 2$ (at the point of equality,
the solution corresponds to $\diffcon$ = 0). When this condition is
met, the solution to equation (\ref{best}) determines the optimum
value of the diffusion constant for which the maximum fraction of
planetary cores survive. 

\begin{figure} 
\figurenum{7}
{\centerline{\epsscale{0.90} \plotone{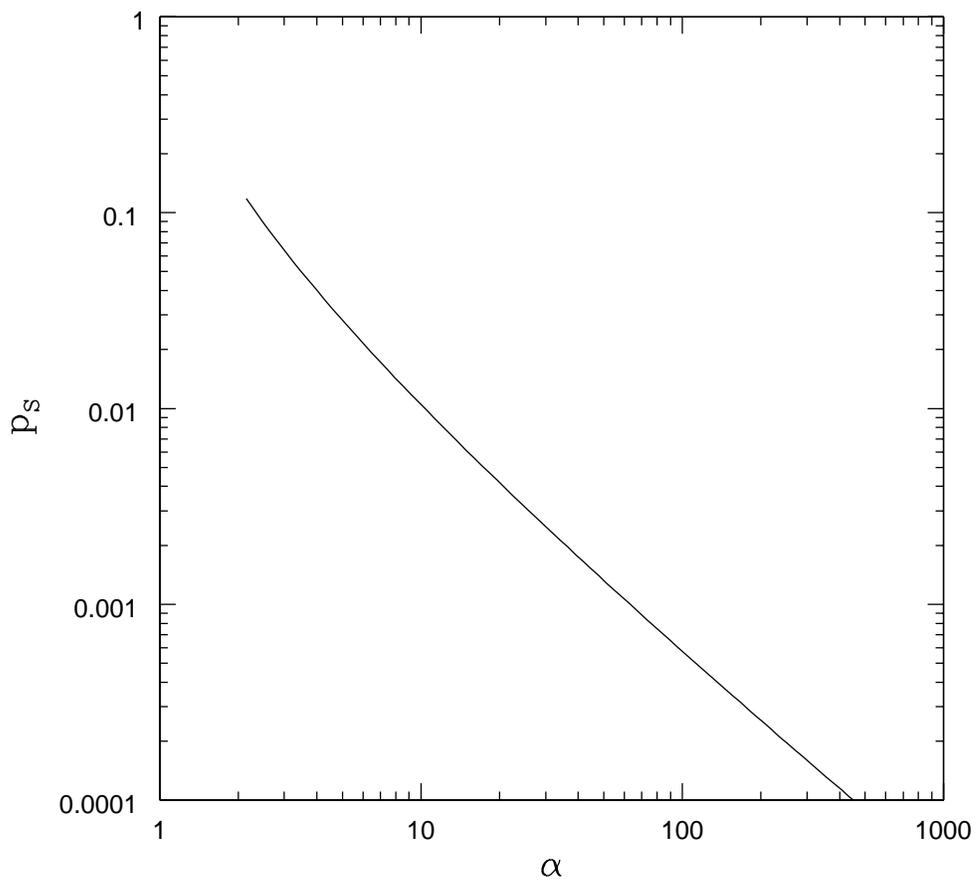} } } 
\figcaption{Fraction $p_S$ of surviving planets as a function of
$\alpha$ = $\gambar t/\sigma$, where the diffusion constant has been
optimized using the self-similar solution of Section 3.4. Notice that
for $\alpha < 2$, no optimizing solution exists; in this regime, 
diffusion acts to reduce the number of surviving planets. For 
$\alpha > 2$, diffusion acts to increase the probability $p_S$ 
of planetary survival. }
\label{fig:alpha} 
\end{figure}

The resulting optimized survival fraction is shown in Figure
\ref{fig:alpha} as a function of the parameter $\alpha = \gambar
t/\sigma$. Notice that for small values of $\alpha < 2$ the
optimization condition (\ref{best}) has no solution.  For the regime
where $\alpha = \gambar t/\sigma < 2$, the fraction of planetary cores
is a decreasing function of the diffusion constant $\diffcon$; in this
regime, the Type I migration has not had time to completely deplete
the planetary population, so that increasing the diffusion constant
leads to loss of planets rather than helping to save them. For the
same choice of parameters used above ($\gambar$ = 0.3 Myr$^{-1}$,
$\sigma$ = 1, and time $t$ = 10 Myr), the optimal survival fraction 
is about $p_S \approx$ 0.064 (compared to the value of $p_S \approx$ 
0.044 obtained previously with $\diffcon$ = 1 Myr$^{-1}$). 
 
\subsection{Long Time Limit} 

The most important outcome of the diffusion process considered herein
is the fraction of surviving planets and their distribution of
positions (given by their angular momentum in this formulation). These
quantities are determined by the solutions to the Fokker-Planck
equation.  Although one can find numerical solutions (see above),
analytic or simplified descriptions can greatly add to our
understanding of the issues.  The analytic treatment developed in
Section 3.3, where the Type I migration torques were turned off, can
be generalized to provide a full solution.  For the full problem,
including the Type I migration term, one can separate variables and
find an analogous series solution.  In this case, however, the spatial
eigenfunctions $F(x)$ are hypergeometric functions (AS), rather than 
Bessel functions, and hence are cumbersome to work with.  Fortunately,
in the long time limit, the problem simplifies greatly. In this
asymptotic limit, only the lowest order term in the expansion survives, 
and the distribution is determined by the solution to the following
eigenvalue problem 
\be
\diffcon {d^2 \over dx^2} \left( x^b F \right) + 
\gamma {d \over dx} \left({F \over x^a} \right) + \lambda_1 F = 0 \, ,
\label{eigenequation} 
\ee
where $\lambda_1$ is the lowest order eigenvalue and $F(x)$ is the
corresponding eigenfunction. Note that we can absorb one of the
parameters. For example, we can divide equation (\ref{eigenequation})
by $\gamma$ and work in terms of a relative diffusion constant
${\widetilde \diffcon} = \diffcon / \gamma$. The scaled eigenvalue 
${\widetilde \lambda} = \lambda_1 / \gamma$ will then be dimensionless.

\begin{figure} 
\figurenum{8}
{\centerline{\epsscale{0.90} \plotone{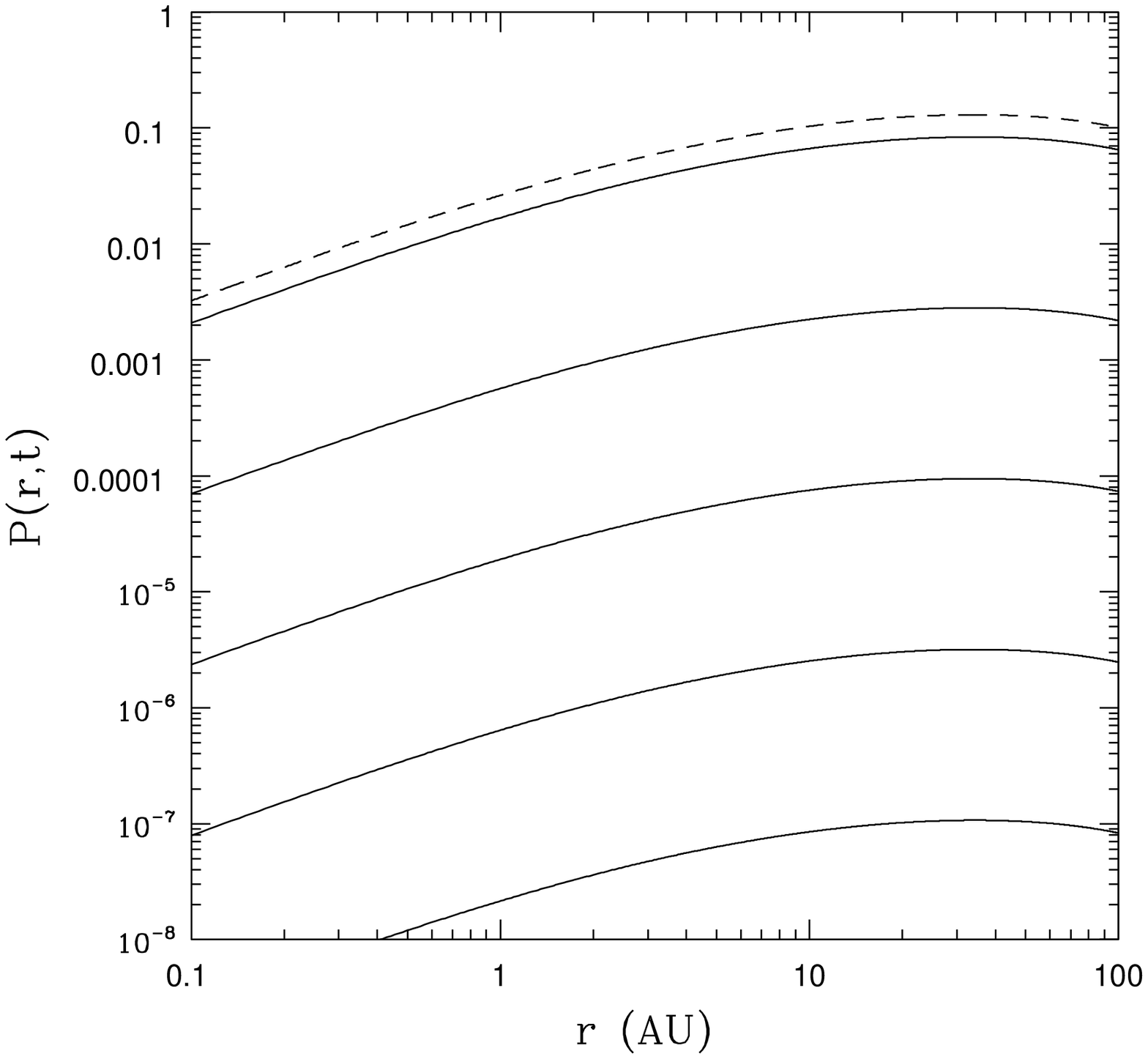} } } 
\figcaption{Distributions $P(r,t)$ in the long time limit.  The solid  
curves show the distributions resulting from numerically integrating
the standard form of the Fokker-Planck equation at five times: 10 Myr
(top curve), 20 Myr, 30 Myr, 40 Myr, and 50 Myr (bottom curve). The
lowest order eigenfunction from equation (\ref{eigenequation}) is
plotted as a dashed curve just above the uppermost solid curve. This 
eigenfunction has almost exactly the same shape as the distributions 
predicted by the Fokker-Planck equation (the eigenfunction must be 
offset from the numerically determined distribution to be visible
in the plot). } 
\label{fig:longtime} 
\end{figure}

Figure \ref{fig:longtime} shows the distributions calculated from our
numerical treatment of the Fokker-Planck equation in the long time
limit.  In this case, the standard form of the Fokker-Planck equation
(with $a$ = 2, $b$ = 1, $\diffcon$ = 1 Myr$^{-1}$, and $\gamma$ = 10
Myr$^{-1}$) was integrated out to 100 Myr. The five solid curves shown
in the figure correspond to times of 10, 20, 30, 40, and 50 Myr, from
top to bottom in the figure.  Notice that the five curves are nearly
parallel to each other and exhibit nearly equal spacing. As a result,
the distributions have reached an asymptotic form, and are decreasing
in amplitude with a well-defined decay rate. The lowest order
eigenfunction calculated from equation (\ref{eigenequation}) is also
shown as a dashed curve, just above the uppermost solid curve.  If
this eigenfunction is plotted with the same normalization as the
distributions resulting from the Fokker-Planck equation, the functions
are indistinguishable. This figure thus demonstrates that the lowest
order eigenfunction provides a good description of the solution in
the long time limit. Furthermore, this limit is reached on a time
scale less than 10 Myr. 

\begin{figure} 
\figurenum{9}
{\centerline{\epsscale{0.90} \plotone{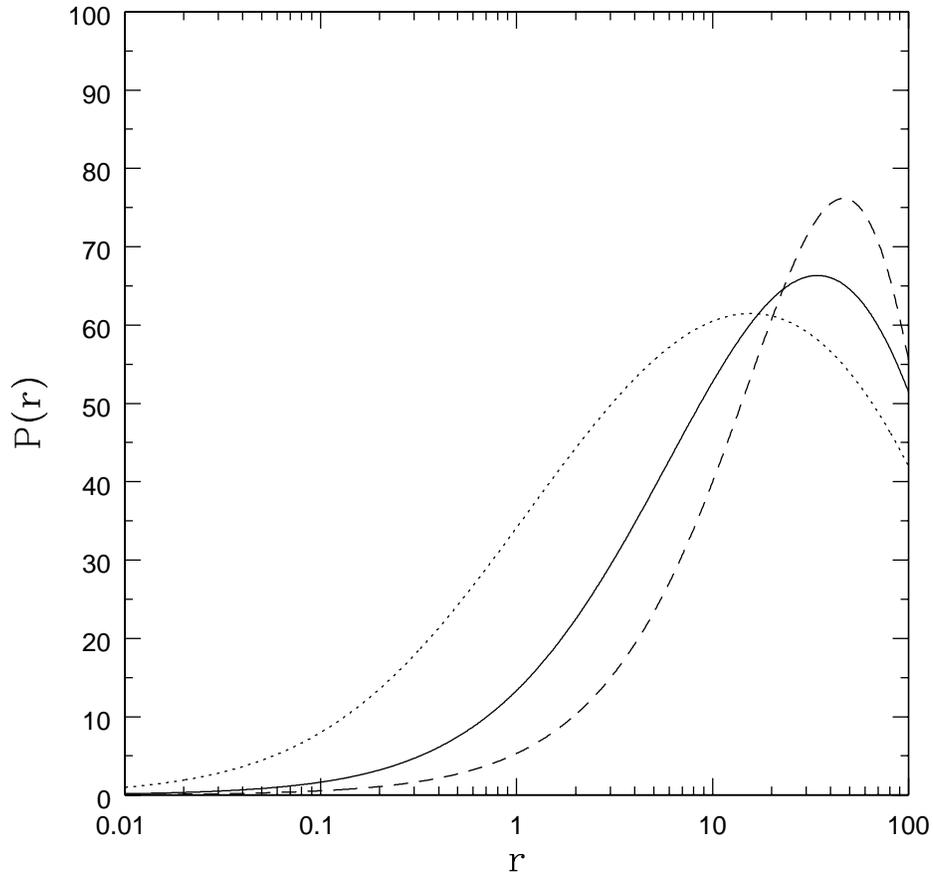} } } 
\figcaption{Eigenfunctions for the lowest order mode solution to the  
Fokker-Planck equation. In the long time limit, these functions
provide the distribution of angular momentum, and hence radial
position, for surviving planetary cores.  The three curves shown here
correspond to a fixed Type I migration parameter $\gamma$ and varying 
values of the diffusion parameter given by $\diffcon/\gamma$ = 0.01 
(dashed curve), 0.1 (solid curve), and 1 (dotted curve). As shown, 
the three eigenfunctions are normalized to the same (arbitrary) value. }
\label{fig:eigenfun} 
\end{figure}

Given that the solutions can be described by the lowest order
eigenfunctions, we can estimate the probable locations for surviving
planetary cores. For given values of the Type I migration parameter
$\gamma$ and the diffusion parameter $\diffcon$, we can find the
eigenvalues and corresponding eigenfunctions for equation
(\ref{eigenequation}). The results are shown in Figure
\ref{fig:eigenfun} for a fixed value of the Type I migration parameter
$\gamma$ and for three values of the diffusion parameter:
$\diffcon/\gamma$ = 0.01, 0.1, and 1; note that only the ratio
$\diffcon/\gamma = \qm$ (see equation [\ref{qdef}]) is needed to
determine the form of the solutions.  As expected, these probability
distributions peak in the outer part of the disk.  As the diffusion
parameter increases, the distributions become wider, and hence have
more support at smaller radii. The three distributions shown in Figure
\ref{fig:eigenfun} are normalized to have the same integrated
value. For larger $\diffcon$ and fixed $\gamma$, however, the survival
probability is a decreasing function of the diffusion parameter in
this regime (see Figure \ref{fig:optim}).

This procedure also specifies the eigenvalues, which in turn determine
the decay rates for planet survival in the long term. For fixed Type I
migration parameter $\gamma$ = 10, and for diffusion parameters
$\diffcon$ = 0.1 , 1, and 10, the lowest order eigenvalues are
$\lambda_1 \approx$ 0.1098, 0.3400, and 2.022, respectively (where all
quantities are in units of Myr$^{-1}$). For diffusion parameter
$\diffcon$ = 1, corresponding to the expected center of parameter
space, this eigenvalue compares favorably with those calculated from
numerical solutions to the Fokker-Planck equation (see Figures 
\ref{fig:ptime} and \ref{fig:lvg}).

\subsection{Time Dependent Torque Parameters} 

Both the Type I migration torque and the stochastic torques due to
turbulent forcing depend on the surface density of the disk.  Since
the disk mass is expected to be a decreasing function of time, the
normalization of the disk surface density will, in general, be time
dependent. To gain some understanding of how this time dependence
affects the migration problem considered herein, we assume that the
disk surface density maintains the same power-law form, but the disk
mass decreases with time.  Toward this end, we introduce a
normalization function $s(t)$ such that the disk mass is given by 
$M_d (t) = M_d (0) s(t)$. Although the form of $s(t)$ is not known, 
observations show that circumstellar disks lose their mass on time
scales of order 3 -- 10 Myr (Haisch et al. 2001, Hern{\'a}ndez et al.
2007, Hillenbrand 2008). More specifically, the observational sample
shows that about half of the stars lose their disks by age $\sim3$
Myr, and that only about $1/e \sim 1/3$ of the disks remain at 5 Myr.
For the sake of definiteness, we use a simple exponential form for
$s(t)$, i.e.,
\be
s(t) = \exp[-t/t_0] \, ,
\label{soft} 
\ee
where we expect the time scale $t_0 = 1 - 10$ Myr. 

Next we note that the Type I migration torque is proportional to the
surface density $\Sigma(r)$, whereas the effective diffusion constant
from the turbulent torques scales like $\Sigma^2$. When the
Fokker-Planck equation is modified to include this time dependence,
it takes the form 
\be
{\partial P \over \partial t} = \gamma s(t) {\partial \over \partial x} 
\left( {P \over x^2} \right)  + \diffcon s^2(t) 
{\partial^2 \over \partial x^2} \left( x P \right) \, ,  
\label{fptimevary} 
\ee 
where we have used the standard radial dependence of the surface
density and temperature (and the standard $x$-dependence of the
torques). Note that $\gamma$ and $\diffcon$ are defined by equation
(\ref{nondim}).

In this formulation, the two terms on the right hand side of the
Fokker-Planck equation (\ref{fptimevary}) display different types of
time dependence. For purposes of illustration, we can consider one
term at a time. For the case in which only one of the torque terms is
operational, we can define a new time variable $\tau$ according to
$d\tau_1 = s(t) dt$ or $d\tau_2 = s^2(t) dt$. With the former
substitution, the Type I migration dynamics becomes the same as that
considered in Section 3.2, with the time $t$ replaced by $\tau_1$.
Similarly, the diffusion dynamics becomes the same as that considered
in Section 3.2, with $t$ replaced by $\tau_2$. With time dependence
surface density, however, the effective time variables $\tau_j (t)$
reach finite values in the limit $t \to \infty$, i.e.,
\be 
\tau_1 (t) = t_0 \left[ 1 - {\rm e}^{-t/t_0} \right] \to t_0 
\qquad {\rm and} \qquad 
\tau_2 (t) = {t_0 \over 2} \left[ 1 - {\rm e}^{-2t/t_0} \right] 
\to {t_0 \over 2} \, . 
\label{teffective} 
\ee 
Thus, the net effect of decreasing disk mass is to limit the operation
of Type I torques to an effective time of $t_0$, and to limit the
operation of diffusion to an effective time of $t_0/2$. On one hand,
the result that decreasing disk mass implies a finite time for torques
to act is expected. Further, the effective time depends on the
function $s(t)$ that describes the time dependence.  On the other
hand, the two types of torques depend on disk mass -- and hence on
$s(t)$ -- in different ways and hence have different effective duty
cycles. For exponential decay in disk mass, Type I migration 
torques act over a time span that is effectively twice as long as 
that of turbulent diffusion. 

\begin{figure} 
\figurenum{10}
{\centerline{\epsscale{0.90} \plotone{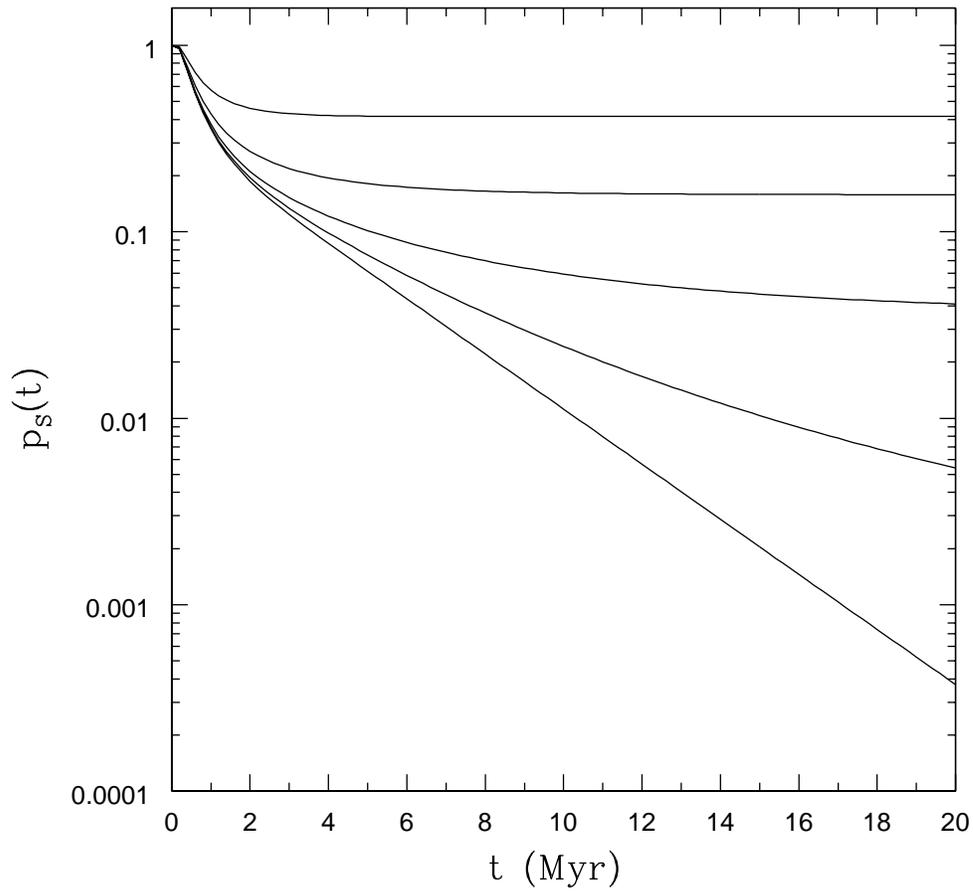} } } 
\figcaption{Survival fraction as a function of time for systems 
where the disk mass decreases with time. The curves shown here
correspond to different exponential time constants for disk mass 
evolution: $t_0$ = 1, 3, 10, 30, and the limit $t_0 \to \infty$ 
(from top to bottom). }
\label{fig:timedep} 
\end{figure}

For a given version of the Fokker-Planck equation, and a given time
dependence $s(t)$ for the surface density and disk mass, we can find
numerical solutions. The result is shown in Figure \ref{fig:timedep}
for the standard choice of power-law disk parameters. The time
dependence of the disk mass has the exponential form given by equation
(\ref{soft}) with different values of the decay time: $t_0$ = 1, 3,
10, 30, and the limit of constant disk mass $t_0 \to \infty$. For each
case, the survival fraction is shown as a function of time. For finite
$t_0$, both types of torques become ineffective over a sufficiently
long span of time, and the survival fraction asymptotically approaches
a constant value. Moreover, for expected values of the disk lifetime, 
these asymptotic values are $p_S \approx 0.04 - 0.16$.

\begin{figure} 
\figurenum{11}
{\centerline{\epsscale{0.90} \plotone{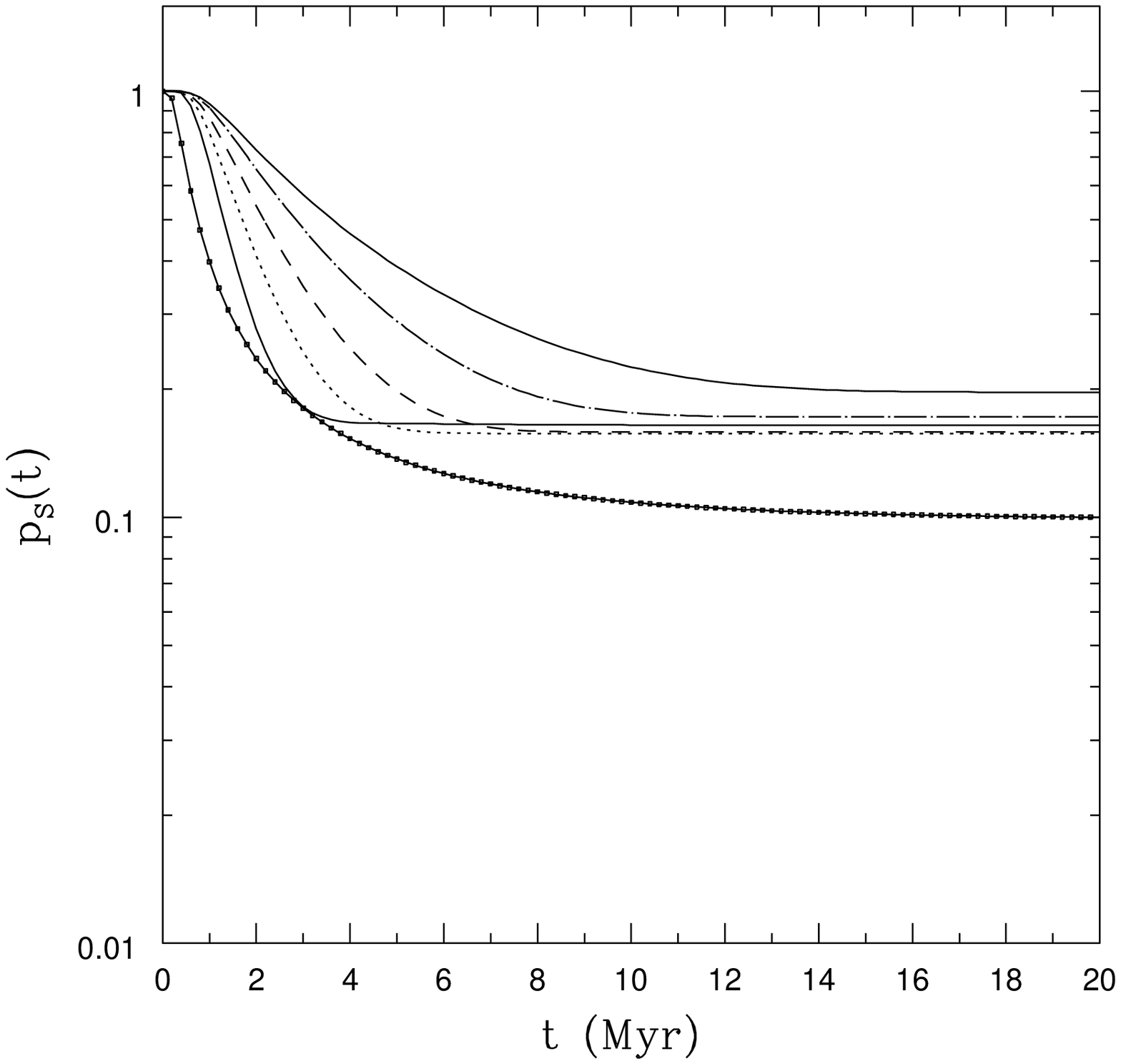} } }  
\figcaption{Survival fraction $p_S$ as a function of time for systems 
where the Type I torque parameter is time dependent and the disk mass
decreases with time. The torque parameter is taken to have the form
given by equation (\ref{gammaform}), where the planetary mass $m_P$ 
increases with time according to $m_P = m_1 (t/{\rm 1 Myr})^3$. 
The disk mass decreases with time constant $t_0$ = 5 Myr.  The curves
shown here correspond to $m_1 / \mearth$ = 1 (lower solid curve), 0.3
(dotted curve), 0.01 (dashed curve), 0.003 (dot-dashed curve), and
0.001 (upper solid curve). The lower curve marked by open squares
shows the result with a constant value $\gamma$ = 10 Myr$^{-1}$. }
\label{fig:timemass} 
\end{figure}

The Type I torque parameter also depends on the mass of the growing
planetary core, and this time dependence can also be included. Here we
present a simple working model to illustrate the type of behavior
introduced by this time dependence. At early times, when the planetary
mass $m_P < 10 \mearth$, the Type I torque parameter depends linearly
on the mass (see equation [\ref{gammadef}]). For larger masses $m_P
\sim 30 - 100 \mearth$, however, the planet clears a gap in the disk, 
and the migration torques become much smaller. We represent this general 
trend by taking the torque parameter $\gamma$ to have the simple form
\be
\gamma = \Gamma (m_P / \mearth) \exp [ - m_P / m_C ] \, . 
\label{gammaform} 
\ee
The function $\gamma (m_P)$ attains its maximum value at $m_P = m_C$.
For the sake of definiteness, here we take $m_C = 10 \mearth$. The
corresponding maximum value is then given by $\gamma$ = $10 \Gamma /
e$. If we use $\Gamma = 10$ Myr$^{-1}$, the maximum value of $\gamma
\approx 30$ Myr$^{-1}$, a typical Type I migration parameter expected
for $m_P \approx 12 \mearth$ (see equation [\ref{gammadef}]). At the
expected gap clearing mass of $m_P \sim 30 \mearth$, the Type I
migration parameter $\gamma \sim 15$ Myr$^{-1}$, and it decreases
rapidly with further increases in $m_P$. 

Next we need to specify the mass of the planetary core as a function
of time.  At relatively small masses, the core grows with accretion
rate ${\dot M} \propto R^2$, where $R$ is the radius of the planet
(e.g., Lissauer \& Stevenson 2007). For constant planetary density,
the mass grows with time according to $m_P \propto t^3$.  At later
times, when the planet is large enough for gravitational focusing to
become important, the accretion rate approaches the form ${\dot M}
\propto R^4$ and the mass increases rapidly.  Once the planetary core
reaches this phase, however, it becomes large enough to clear a gap
and the Type I torques are significantly less important. We thus
concentrate on the early phase, and hence allow the planetary mass to
grow according to $m_P = m_1 (t/{\rm 1 Myr})^3$. The parameter $m_1$
depends on the surface density of solids in the disk and the radius of
core formation.  Here we take this mass scale to lie in the range
$m_1/\mearth$ = 0.01 -- 1.  For this range, the corresponding time
required for a growing planetary core to reach the threshold value of
$m_P$ = 10 $\mearth$ is $t \sim 2 - 10$ Myr.

Using the above time dependence for the Type I torque parameter, the
Fokker-Planck equation can be integrated as before, also including the
decrease in disk surface density through the function $s(t)$. The
result is shown in Figure \ref{fig:timemass} for a disk evolutionary
timescale of $t_0$ = 5 Myr, and for $m_1$ = 0.01 -- 1 $\mearth$.  The
curve for $\gamma$ = 10 Myr$^{-1}$ = {\sl constant} is also shown (for
the same timescale $t_0$). In the scenario with time dependent
planetary mass, the Type I torque parameter is smaller than our
assumed constant value at early times, but larger at later times. To
leading order, the time dependence tends to cancel out. Since
planetary core masses grow rapidly, however, the systems spend more 
time with lower torque parameter values, so that the inclusion of this
time dependence allows more planetary cores to survive (see Figure
\ref{fig:timemass}). We can understand this result by defining an
effective duty cycle $\tau_3$ for the Type I torques, analogous to
those in equation (\ref{teffective}), by including both the time
evolution of the planetary mass and the disk surface density. For the
parameters used here, this time scale lies in the range $\tau_3
\approx 4.2 - 5.6$ Myr. These timescales are close to that for disk
evolution only, $\tau_1 = t_0$ = 5 Myr, indicating that the smaller
values of $\gamma$ at early times nearly cancel the larger values at
later times. This (approximate) cancellation is reflected in the 
survival fractions, which are confined to the range $p_S \approx$ 
0.1 -- 0.2 for the cases shown in Figure \ref{fig:timemass}. Of 
course, disk systems can display a wide range of parameters, so 
that smaller survival fractions can also be realized. 

\section{CONCLUSION}

This work reinforces and extends results obtained in previous studies
(LSA, NP, JGM): Turbulence transforms Type I migration from a steady
inward progression into a diffusive process.  Turbulence thus allows
some fraction of the population of planetary cores to survive beyond
the Type I migration timescale.  However, the outcome of any
particular migration episode is uncertain because of extreme
sensitivity to initial conditions --- due to chaos --- so that the
results must be described in terms of probability distributions (see
Figures \ref{fig:pdist}, \ref{fig:longtime}, \ref{fig:eigenfun}, and
\ref{fig:iterate}). This survival problem, where steady inward
migration is coupled to stochastic behavior, and where the torques
associated with both effects can vary with time, allows for a rich
diversity of behavior.  A more specific description of our results is
given below (Section 5.1) along with a discussion of their
implications and limitations (Section 5.2).

\subsection{Summary of Results} 

Stochastically driven diffusion, due to turbulent torques, can act to
save planetary cores from accretion due to Type I migration.  For
torque strengths near the center of the expected range of parameter
space (specifically, $\gamma$ = 10 Myr$^{-1}$, $\diffcon$ = 1
Myr$^{-1}$, and active disk lifetime $t$ = 3 Myr), and for planets
starting near $r$ = 10 AU, the survival fraction $p_S \approx 0.1$
(see Figures \ref{fig:ptime}, \ref{fig:ptimebeta}, and \ref{fig:optim}).  
Note that this ``lifetime'' can be the time required for the planetary
core to reach the threshold required for gap clearing, so that
migration slows down.  For longer timescales, the fraction of
surviving bodies is much smaller. For the same torque parameters, the
survival fraction $p_S \approx 0.01$ at $t$ = 10 Myr and $p_S \approx$
0.0004 at $t$ = 20 Myr. Keep in mind that these survival fractions are
modified when the torque parameters exhibit time dependence (see below). 

The outer boundary condition in the disk plays an important role in
determining the fraction of surviving planets. A finite disk edge
causes the fraction of surviving planets $p_S(t)$ to experience
exponential decay (see Figure \ref{fig:ptime}, Sections 3.3 and 4.3),
whereas a disk with infinite extent displays power-law decay (see
Section 3.4 and JGM).  In most cases of interest, the expected disk
outer radius ($\rdisk \sim 30 - 100$ AU) is small enough that planets
can diffuse to the outer boundary during the active disk lifetime, so
that edge effects are important and exponential decay is realized. 
Typical decay rates lie in the range $\lambda$ = 0.1 -- 0.5 Myr$^{-1}$, 
and are found from both analytic calculations (Sections 3.3 and 4.3) 
and numerical simulations (Figures \ref{fig:ptime} and \ref{fig:lvg}).

The probability of planet survival is sensitive to the initial
conditions.  The most favorable locations for forming planetary cores
lie just outside the snow-line in circumstellar disks, i.e., in the
radial range 5 -- 10 AU. For typical torque parameters, this regime 
also marks the boundary between the outer disk, where turbulent
torques dominate, and the inner disk, where inward Type I migration
torques dominate. As a result, planetary cores starting their
migration within this annulus are particularly sensitive to the
specifics of their evolution. Planets forming at somewhat larger radii
are much more likely to survive, whereas planets that form at smaller
radii have little chance of survival (Figure \ref{fig:rstart}).

For a given value of the Type I inward migration torque amplitude,
there exists an optimum value of the diffusion constant that leads to
the maximum number of surviving planets. This extremum depends on the
effective disk lifetime. For sufficiently short lifetimes (shorter
than the nominal Type I migration time), diffusion acts to 
{\it reduce} the fraction of surviving planets and the optimum value
of the diffusion constant is zero. For longer disk lifetimes,
diffusion acts to save planets, and a maximum develops in the survival
curve (see Figure \ref{fig:optim}). The optimum value of the diffusion
constant corresponds to an optimum level of turbulence.  Furthermore,
this optimum level of turbulence is relatively near that found in
previous MHD simulations (LSA, NP, Nelson 2005). The existence of an
optimal value of the diffusion constant can be derived analytically
using the self-similar limiting form of the problem (see Section 4.2,
Figure \ref{fig:alpha}, and equations [\ref{selfsimint}, \ref{best}]).

In the long time limit, the distributions of angular momenta for
surviving planets approach a well-defined form (see Figure
\ref{fig:longtime}), with the amplitude (normalization) decreasing at
a well-defined decay rate.  The form of this asymptotic distribution
is given by the lowest order eigenfunction of the spatial part of the
Fokker-Planck equation (see Figure \ref{fig:eigenfun}), and the decay
rate is given by the corresponding eigenvalue (see equation
[\ref{eigenequation}]). The distribution of surviving planets peaks in
the outer disk and provides the initial conditions for the later
stages of planetary growth.

The time dependence of the disk mass and surface density leads to
corresponding time dependence in the torque parameters, and can be
incorporated into this formulation of the diffusion problem (Section
4.4). Because the two types of torques depend on different powers of
the surface density, the effective duty cycle of the Type I migration
torque is longer than that due to turbulence. When the time dependence
of the disk surface density is included, the survival probability of
planets approaches a well-defined asymptotic value (Figure
\ref{fig:timedep}) that depends on the disk evolutionary timescale
$t_0$ (equation [\ref{soft}]).  For standard torque parameters and
$t_0$ = 3 Myr (5 Myr) --- consistent with observed disk timescales
(Hern{\'a}ndez et al. 2007) --- the survival fraction has values $p_S
\approx$ 0.16 (0.10). The Type I torque parameter also depends on the
mass of the planetary core, which grows with time. The inward
migration torques are thus smaller than average at early times and
larger at later times.  When this time dependence is included, the net
survival probability is increased by a modest amount (see Figure
\ref{fig:timemass}), with typical values $p_S$ = 0.1 -- 0.2.

For completeness, we have developed an alternate description of the
dynamics using an iterative map formalism (given in the Appendix).
Although this treatment gives the same results as the Fokker-Planck
equation for the same input physics, an iterative map can be useful in
several ways: The mapping provides another way to derive --- and hence
understand --- the way in which a finite disk edge enforces an
exponentially decreasing survival probability (Section A.3).  The
Fokker-Planck treatment is limited to small diffusion steps, whereas
the iterative map can accommodate large fluctuations. Since boundary
conditions are implemented in different ways in the two treatments,
some boundary conditions are easier to model with the mapping
approach. Finally, the iterative map can easily be generalized to
include eccentricity variations and other complications.  While the
iterative map approach is flexible and instructive, it is very
computationally intensive: To obtain each of the distributions shown
in Figure \ref{fig:iterate}, we needed to perform 100,000 random-walk
experiments. This large number is required, in part, because the
survival rate is low. In any case, this finding underscores the
necessity of using complementary methods such as the Fokker-Planck
equation.

\subsection{Discussion} 

One of the interesting results of this study is the complicated nature
of the Type I migration epoch. In particular, the formation and
survival of planetary cores involves a series of compromises: [A] In
disks with typical properties, Type I torques dominate in the inner
disks where $r < 10$ AU, and stochastic torques dominate in the outer
disk ($r > 10$ AU).  The planetary cores are most easily formed just
outside the snow-line, near 5 AU for solar-type stars and typical
disks; core formation at larger radii is increasingly difficult (as
$r$ increases) due to the slower orbit time.  As shown here, however,
the survival of these cores is enhanced if they start migration at
larger radii (Figure \ref{fig:rstart}).  [B] Next we find that
although turbulence allows planetary cores to survive in spite of Type
I migration (Figures \ref{fig:pdist}, \ref{fig:ptime}, and
\ref{fig:ptimebeta}), the survival fraction decreases if the diffusion
constant becomes too large (Figure \ref{fig:optim}).  [C] We also find
that decreasing the disk surface density with time allows for more
planetary cores to survive (Figures \ref{fig:timedep} and
\ref{fig:timemass}); if the surface density decreases too quickly,
however, the disk will not have enough gas left to make giant planets.
[D] Similarly, the surviving cores are most likely to reside in the
outer disk, near $\sim30$ AU (Figure \ref{fig:eigenfun}); however,
planet formation proceeds much more slowly at large radii and the
outer portion of the disk is most susceptible to mass loss through
photoevaporation (e.g., Adams et al. 2004).  Because of these
compromises, the survival of planetary cores depends on the interplay
between a large number of ingredients, and the Type I migration epoch
results in a wide distribution of possible outcomes.  These
complications, in turn, imply that the resulting planetary systems
will display a great deal of diversity.

Although this paper generalizes previous work, a number of additional
issues remain to be addressed. We first note that the parameter space
for studying the Type I migration problem is huge: In addition to the
magnitude of the torque parameters, and their variations with radius,
the time dependence of the disk surface density and the planets also
play an important role. Next, the true nature of turbulence in
circumstellar remains under study, so that its effects on planet
migration could vary from system to system and could otherwise alter
the assumptions used herein.  One important issue is that the
numerical simulations that predict turbulence are not fully converged,
so that changes in numerical predictions are possible (e.g., Fromang
\& Papaloizou 2007).  The formulation presented here separates the 
Type I migration torque from the stochastic turbulent torques and
``derives'' their amplitudes independently. In practice, however, the
presence of turbulence is likely to affect the structure of the disk
near the forming planet and can thus alter the Type I torques (e.g.,
Papaloizou et al. 2007).  Fortunately, our formulation of the
migration problem is sufficiently general to address these issues. 
If, for example, turbulence alters the size of the Type I migration 
torque, or even if it produces a net torque with nonzero mean, this
effect can be incorporated by using the proper value of $\gamma$.
Another unresolved issue is the correlation time of the turbulence
(taken here to be one orbit time). This issue affects the value of the
diffusion constant $\diffcon$.

Another issue that affects the survival of planetary cores is the
possible presence of ``dead zones'', i.e., regions in the disk where
MRI is not active because of insufficient ionization (Gammie 1996).
In these zones, turbulence is absent and hence the diffusion constant
vanishes.  Since Type I torques continue to operate in these regions,
planetary cores migrate inward and can be lost.  Although the
structure and radial extent of dead zones in disks are not fully
understood, these zones are expected in the annulus from about 0.3 to
3 AU.  In the extreme case, the outer edge of the dead zone ($\sim3$
AU) would provide the effective inner boundary for the diffusion
problem addressed in this paper --- all planets that reach this
location would quickly be swept inward by Type I torques and
eventually accreted by the star.  However, this picture contains many
complications: The outer (top/bottom) layers of the disk remain
ionized, and hence turbulent. These regions provide some (highly
reduced) torques (Oishi et al. 2007), and allow for turbulent mixing
that can enliven the dead zones (Ilgner \& Nelson 2008, Inutsuka \&
Takayoshi 2005). In addition, the lower viscosity in the dead zone can
allow the planet to open a gap at lower masses and thereby reduce its
inward speed (e.g., Matsumura et al. 2007). These issues render the
migration scenario complex, and should be addressed in future work.

Finally, we note that this paper only addresses the survival of
planetary cores. Many additional steps are required to produce fully
formed giant planets. After the planetary cores reach a sufficiently
large mass (30 -- 100 $\mearth$), they clear gaps in the disks and
migrate more slowly. This study shows that the planetary cores that
survive the embedded phase of migration will reside in the outer disk
(Figure \ref{fig:eigenfun}). If the planets did not migrate after
clearing gaps, the results of this model would predict many more giant
planets in wide orbits ($a$ = 10 -- 30 AU) compared with those in
close orbits ($a \approx 0.1$ AU); the outer planets would be more
abundant by a factor of 6 to 100, depending on the values of the
torque parameters (see Figure \ref{fig:eigenfun}).  However, the
surviving cores will move inward through Type II migration as they
continue to grow. This later migration phase is not calculated herein,
but it will act to move the distribution of semi-major axes inward, and
should be considered in future work.

The basic issue addressed in this paper is that Type I migration tends
to move planetary cores inward too rapidly, before they can clear gaps
and before they can grow into giant planets. Building on previous work
(LSA, NP, JGM), we have explored a solution to this Type I migration
problem where the planetary cores experience a random walk due to
turbulent perturbations. Although this solution is successful in many
ways, other physical processes can contribute.  If the planetary orbit
is eccentric, for example, the Type I torques are weaker (Papaloizou
\& Larwood 2000); migration can thus be slowed down if some process
can maintain orbital eccentricities.  Similarly, the torques are
weaker if the disk itself maintains global (non-axisymmetric)
distortions (Papaloizou 2002).  Another contributing factor is the
detailed structure of the disk, which can depart from the power-law
forms considered here.  Opacity transitions affect the disk structure
and hence the migration rates (Menou \& Goodman 2004); for
sufficiently high opacities, the migration can even be directed
outwards (Paardekooper \& Mellema 2006). Strong magnetic fields can
dominate over Type I torques (Terquem 2003), moving planets both
inward and outward.  Finally, the inner disk can be truncated by
magnetic effects (Shu et al.  2007) so that planetary cores cannot
migrate all the way to the stellar surface.  In closing, the challenge
left for the future is to determine how all of these processes --- and
others --- work together to extend the time required for Type I
migration and thereby allow giant planets to form.


\acknowledgements 

We thank Nuria Calvet, Greg Laughlin, and Daniel Lecoanet for useful
discussions.  This work was supported in part by the Michigan Center
for Theoretical Physics. FCA is supported by NASA through the Origins
of Solar Systems Program via grant NNX07AP17G.  AMB is supported by
the NSF through grant DMS-604307. In addition, AMB and FCA are jointly
supported by Grant Number DMS-0806756 from the NSF Division of Applied
Mathematics.

\appendix 
\section{AN ITERATIVE MAPPING SCHEME FOR MIGRATION} 

As an alternative to the Fokker-Planck treatment presented in the
text, this Appendix develops an simple iterative mapping approach.
The evolution of planetary angular momentum evolution can be described
by an iterative map that includes both the Type I inward migration and
the stochastic changes due to turbulent forcing. The map can thus be
written in the form
\be 
j_{k+1} = 
\left[1 - \left( {\Delta j \over j} \right)_1 \right]_k
\left[1 + \left( {\Delta j \over j} \right)_T \right]_k j_k \, ,
\ee
where the subscript labels the step number. Note that the angular
momentum increments occur over the time scale $\tturb$ on which 
the turbulent fluctuations are independent (roughly an orbit time). 
As a result, we must include a second map to track the time, i.e., 
\be
t_{k+1} = t_k + (\tturb)_k = t_k + {2 \pi f_\alpha \over \Omega_k} \, . 
\ee 
The full map can be written in terms of an ordered product. The 
angular momentum at orbit number $N$ becomes 
\be 
j_N = j_0 \prod_{k=1}^N 
\left[1 - \left( {\Delta j \over j} \right)_1 \right]_k
\left[1 + \left( {\Delta j \over j} \right)_T \right]_k \, , 
\ee 
where $j_0$ is the starting value. Note that the factors, in general,
depend on angular momentum and are evaluated at the previous step. As
a result, the order of the product matters. Although the random
parameter $\xi$ that determines the realization of the turbulent
torque is independently distributed, the angular momentum increments
due to turbulence are not.

\subsection{An Aside on Mapping Approximations} 

In the treatment given above, we made the approximation such that the
angular momentum increments are small per orbit. In particular, we
have multiplied the torque by the orbit time scale $2\pi f_\alpha/\Omega$ 
instead of integrating over the same time interval. For power-law
disks, one can easily perform the integration and obtain more accurate
formulae. In practical terms, however, the uncertainties in the
turbulent forcing are larger than the accuracy gained. In order to
understand the relationship between the iterative map and the
Fokker-Planck treatments, however, we need the integrated result. For
the benchmark case where $[(\Delta j)/j]$ is constant, we thus obtain
\be
dj = - T_1 dt = - [T_1]_k \left( {j \over j_k} \right)^{-2} 
{2 \pi \over \Omega} dn = - [T_1]_k {2 \pi \over \Omega_k}
{j \over j_k} dn \, , 
\ee
where the subscript $k$ denotes that the quantities are to be
evaluated at the beginning of the $kth$ time interval. Here, $dn$ is
the increment of the number of orbits, so we need to integrate from
$n=0$ to $n=f_\alpha$. After integrating, the change in angular
momentum over the time scale $\tturbk$ (for the current radial 
location) takes the form
\be
1 - \left( {\Delta j \over j} \right)_1 = 
\exp \left\{ - {[T_1]_k \over j_k } \, \, \tturbk \right\} \, . 
\ee
Note that the product of many such factors takes the form 
\be
\Pi^{(N)} = \prod_{k=1}^N 
\left[ 1 - \left( {\Delta j \over j} \right)_1 \right]_k =
\prod_{k=1}^N \exp \left\{ - {[T_1]_k \over j_k } 
\, \, \tturbk \right\} = \exp \left\{ - \, \sum_{k=1}^N \, 
{[T_1]_k \, \tturbk \over j_k } \right\} \, . 
\ee
For this case, $[T_1] \propto j^{-2}$ and $\tturb \propto \Omega^{-1}$
$\propto j^3$ so that $[T_1] \tturb/j$ is a constant (the same for
each cycle). As a result, we can write the product in the form 
\be
\Pi^{(N)} = \exp \left\{ - \, \sum_{k=1}^N \, 
{[T_1]_0 \, \tturbk \over j_0 } \right\} =
\exp \left\{ - {[T_1]_0 \, \tturbk \over j_0 } N \right\} = 
\left[ \exp \left\{ - {[T_1]_0 \, \tturbk \over j_0 } \right\} 
\right]^N \, , 
\ee
where the subscript zero denotes that the quantities are to be 
evaluated at the beginning of the migration epoch ($t$ = 0). 

\subsection{Example} 

As a working example, we consider the standard disk where $q$ = 1/2
and $p$ = 3/2. In this case, the relative angular momentum changes due
to Type I migration are independent of $j$, i.e., the planet loses a
fixed fraction of angular momentum per orbit (or per time interval
$\tturb$). The effect of Type I migration on the planet is just a
constant factor $F_1$ in the iterative map. Here we take
\be
\left( {\Delta j \over j} \right)_1 = 10^{-5} 
\qquad \Rightarrow \qquad F_1 = \exp \left[ - 10^{-5} \right] \, .
\ee 
For this same disk model, the relative angular momentum perturbations 
due to turbulent fluctuations are linear in $j$, and the iterative 
map takes the form 
\be
j_{n+1} = F_1 \left[1 + A \xi (j_n/j_0) \right]_n j_n \, ,
\ee
where $\xi$ is a random variable and $A$ sets the amplitude.  Here we
take $\xi$ to follow a gaussian distribution with zero mean and unit
variance.  The amplitude is set to $A = 10^{-3}$, which corresponds
to our standard value $\beta$ = 1 in the Fokker-Planck equation (see
Section 4). The scale $j_0$ is the angular momentum for a circular
orbit at $a$ = 1 AU around a solar type star ($M_\ast$ = 1.0 $M_\odot$).

\begin{figure} 
\figurenum{12}
{\centerline{\epsscale{0.90} \plotone{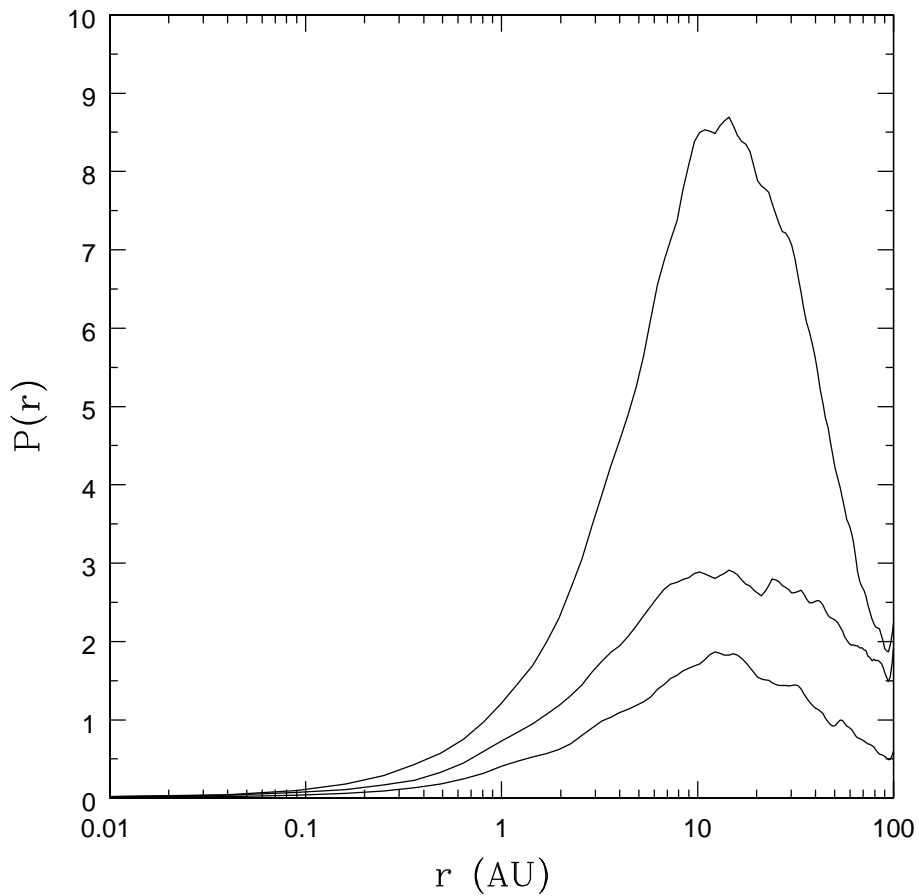} } } 
\figcaption{Distributions of migrating planetary cores at
three times: 1 Myr, 3 Myr, and 5 My (from top to bottom).
These results provide an example of the iterative mapping 
scheme developed in this Appendix. Here, the initial 
condition is taken to be a narrow gaussian distribution
centered on $x = \sqrt{30}$, i.e., the angular momentum
appropriate for a circular orbit at radius $r$ = 30 AU. } 
\label{fig:iterate} 
\end{figure}

The starting radii are taken to be distributed in a narrow gaussian
centered on $x$ = $\sqrt{30}$, corresponding to the angular momentum
appropriate for a circular orbit at $r$ = 30 AU.  The resulting
distributions of radial locations are shown in Figure
\ref{fig:iterate} for times $t$ = 1, 3, and 5 Myr. These distributions
are both qualitatively and quantitatively like those produced by
solutions to the Fokker-Planck equation (see the main text).  Compared
to the Fokker-Planck solutions, these distributions have a slightly
smaller height near the outer boundary. This discrepancy is due to the
different ways in which the outer boundary condition is implemented in
the two methods. In the Fokker-Planck treatment, we use a standard
zero-flux condition at the disk edge. For this iterative map approach,
the migrating planet is not allowed to cross the radius corresponding
to the disk edge. This latter boundary condition is thus implemented
in a ``one planet at a time'' manner and does not exactly produce the
smooth (zero-derivative) solutions of the Fokker-Planck equation.

\subsection{Heuristic Argument for Exponential Decay} 

We can use this iterative map formalism to show that the number of
surviving planetary cores is a decaying exponential function. This
argument applies when the disk has a well-defined outer edge.

We consider only late times, when most planets would be swept into the
star by Type I migration in the absence of diffusion. In this regime,
most of the surviving planets will be piled up in the vicinity of the
outer disk edge. The change in angular momentum due to turbulent
torques, which in general depend on the planet's location, can be
simplified by evaluating the torque amplitude at a constant value near
the disk edge. The change in angular momentum due to Type I migration
is already (in the standard case) a constant. As a result, the net
effect of one cycle of the iterative map is to change the angular
momentum by the factor
\be
{\cal F} = {\cal F}_1 {\cal F}_T = 
\left[ 1 - \left( {\Delta j \over j} \right)_1 \right] 
\left[ 1 + \left( {\Delta j \over j} \right)_T \xi \right] \, , 
\ee
where $\xi$ is a random variable of zero mean and unit variance, 
and the other factors are now constant. After $N$ iterations,  
the accumulated angular momentum can be written as the product
\be
j_{(N)} = j_0 \, {\cal F}_1^N \, \prod_{k=1}^N 
\left[ 1 + \left( {\Delta j \over j} \right)_T \xi_k \right] \, , 
\ee
which can be rewritten in the more convenient form 
\be
\ln \left[ j_{(N)} / j_0 \right] = N \ln {\cal F}_1 + 
\sum_{k=1}^N \ln \left[ 1 + 
\left( {\Delta j \over j} \right)_T \xi_k \right] \, . 
\ee
Working to leading order, we simplify the sum so that 
the angular momentum variable takes the form 
\be
\ln \left[ j_{(N)} / j_0 \right] = N \ln {\cal F}_1 + 
\left( {\Delta j \over j} \right)_T \, \sum_{k=1}^N \xi_k \, , 
\label{ransum} 
\ee
which is correct to the same order as the Fokker-Planck treatment (see
below). The final sum in equation (\ref{ransum}) is the sum of random
variables. In the limit of large $N$, the long time limit, the
composite variable $\zeta_N = \sum \xi_k$ will have a distribution
that approaches a normal form (due to the Central Limit Theorem, e.g.,
Richtmyer 1978). Further, since the individual variables $\xi_k$ have
unit variance, the composite variable $\zeta_N$ has variance
$\sigma_\zeta^2 = N$. 

Survival of the planetary core requires that the angular momentum
remain larger than that of the star, i.e.,
\be
j_{(N)} > j_\ast \equiv j_0 \, {\cal F}_1^K \, , 
\ee
where the second equality defines $K$, the number of steps required 
for Type I migration to reduce the angular momentum of the starting 
state $j_0$ to that of the stellar surface $j_\ast$. Combining the 
above results implies the following requirement for planetary survival
\be
\zeta_N > { (N - K) \ln {\cal F}_1^{-1} \over 
\left( \Delta j / j \right)_T \, } \approx (N - K) 
{ (\Delta j / j)_1 \over (\Delta j / j)_T \, } \equiv 
\zeta_\ast \, .   
\ee
The probability of planetary survival $p_S$ is thus given by 
the integral 
\be
p_S (N) = A \int_{\zeta_\ast}^\infty \exp[ - \zeta^2 / 2N] d\zeta \, , 
\label{survivint} 
\ee
where $A$ is a normalization constant. Note that the planetary cores
do not necessarily have a gaussian distribution in their initial
state, so that the constant $A$ can be less than that corresponding to
the standard normalization at $t$ = 0.  In the regime of interest, at
late times when $N$ is large, the integral in equation (\ref{survivint}) 
can be evaluated asymptotically to obtain 
\be
p_S (N) = A \, {N \over \zeta_\ast } \, \exp[-\zeta_\ast^2/2N] \, 
\left[ 1 - {N \over \zeta_\ast^2} + {3 N^2 \over \zeta_\ast^4} + 
\dots \right] \, . 
\ee
In the extreme limit $N \gg K$, the survival probability can be 
written in the form 
\be
p_S (N) = {A \over {\cal R} } \, \exp[- ({\cal R}^2/2) N ] \, , 
\label{exdecay} 
\ee
where ${\cal R} \equiv (\Delta j/j)_1 / (\Delta j/j)_T$.  Since Type I
migration dominates in the inner disk, but the torques increase their
amplitude relative to Type I torques as the radius increases, we
expect the ratio ${\cal R}$ to be order unity in the outer disk. In
addition, since the parameter $N$ counts orbits, but the orbits in
question are those near the outer disk edge, the parameter $N$ is
proportional to time. Thus, the above result shows that the survival
probability decays exponentially with time.

The decay rate is overestimated in the above analysis because we have
taken the limit $N \gg K$. In practice, orbits will decay due to Type
I torques in $K \sim 10^5$ orbits, typically a few Myr, so that $N$
will be comparable to (but still larger than) $K$. This correction
does not change the result that the fraction of surviving planets
decays exponentially, but it does lower the decay rate. Operationally,
$N$ should be replaced by $(N-K)^2/N$ in the argument of the
exponential in equation (\ref{exdecay}).


\end{document}